\documentclass[usenatbib,dvipsnames,longauth]{aa}
\usepackage{subfig}
\usepackage[T1]{fontenc}
\usepackage{ae,aecompl}
\usepackage{graphicx} 
\usepackage{amsmath}  
\usepackage{amssymb}  
\usepackage{lscape}
\usepackage{rotating}
\usepackage{comment}
\usepackage{hyperref}
\usepackage{txfonts}
\usepackage{microtype}
\usepackage{xcolor} 
\usepackage{soul} 
\usepackage{enumitem}
\usepackage{booktabs}
\usepackage{placeins}
\usepackage[normalem]{ulem} 
\providecommand{\LEt}[1]{} 
\providecommand{\Let}[1]{} 
\providecommand{\simsaple}{\simsample} 

\newcommand{\planck}{{\it Planck}}
\newcommand{\chxmate}{CHEX--MATE}
\newcommand{\simsample}{The300}
\newcommand{\truene}{$n_{\rm e \; 3D}$}
\newcommand{\truemgas}{$M_{\rm gas \;  3D}$}
\newcommand{\truemgasv}{$M_{\rm gas \; 3D,500}$}
\newcommand{\truedlog}{$dlog \; n_{\rm e \; 3D}$}
\newcommand{\Rv}{$\rm R_{500}$}
\newcommand{\xmm}{XMM-\textit{Newton}}
\newcommand{\nemedian}{$\widetilde{ n_{\rm e}}$ } 
\newcommand{\nemean}{$\langle n_{\rm e} \rangle$}

\def\threehun{{\sc The Three Hundred }}
\def\threehunnumbers{{\sc The300}}

\def\lesssim{\mathrel{\hbox{\rlap{\hbox{\lower4pt\hbox{$\sim$}}}\hbox{$<$}}}}

\usepackage{lipsum}

\hypersetup{colorlinks,linkcolor={MidnightBlue},citecolor={MidnightBlue}}

\begin{document}
\title{CHEX-MATE: Density profile reconstruction in  galaxy clusters}

\author{I. Bartalucci \inst{\ref{tov},\ref{iasfmi},\ref{tovinfn}}, 
        G. W. Pratt \inst{\ref{cea}} ,
        E. Rasia \inst{\ref{inaf-trieste},\ref{ifpu-trieste},\ref{michigan}} ,
        M. Arnaud \inst{\ref{cea}} ,
        R.T. Duffy \inst{\ref{cea}} , 
        Y. E. Bahar \inst{\ref{inaf-oas}} ,
        M. Balboni \inst{\ref{unibo},\ref{iasfmi}} ,
        E. Baraldi \inst{\ref{cea}} ,
        R. M. Batalha \inst{\ref{obs-brasil}} ,
        A. Bonafede \inst{\ref{unibo},\ref{ira}} ,
        H. Bourdin \inst{\ref{tov},\ref{tovinfn}} ,
        F. De Luca \inst{\ref{tov},\ref{tovinfn}} , 
        M. De Petris \inst{\ref{sapienza}} ,  
        S. Ettori \inst{\ref{inaf-oas},\ref{infn-bo}} ,
        A. Ferragamo \inst{\ref{uninapoli},\ref{sapienza}} ,
        M. Gaspari \inst{\ref{unimodena}} ,
        F. Gastaldello \inst{\ref{iasfmi}} ,
        M. Gitti \inst{\ref{unibo},\ref{ira}} ,
        A. Iqbal \inst{\ref{lille}} ,
        M. Lepore \inst{\ref{cea},\ref{arceteri}} ,
        S. T. Kay \inst{\ref{manchester}} , 
        L. Lovisari \inst{\ref{iasfmi},\ref{cfa}} ,
        P. Mazzotta \inst{\ref{tov},\ref{tovinfn}} ,
        B. J. Maughan \inst{\ref{bristol}}     ,
        F. Oppizzi \inst{\ref{infn-genova}}   ,
        E. Pointecouteau \inst{\ref{irap}}   ,          
        M. Rossetti \inst{\ref{iasfmi}}  ,
        J. Sayers \inst{\ref{caltech}}   ,
        M. Sereno \inst{\ref{inaf-oas},\ref{infn-bo}}}
        
   \institute{Dipartimento di Fisica, Università degli studi di Roma Tor Vergata, Via della Ricerca Scientifica 1, I-00133 Roma, Italy \label{tov} \and
INAF/IASF-Milano, Via A. Corti 12, 20133 Milano, Italy \label{iasfmi} \and
INFN, Sezione di Roma ‘Tor Vergata’, Via della Ricerca Scientifica, 1, 00133, Roma, Italy \label{tovinfn} \and
Universit'e Paris-Saclay, Universit'e Paris Cit'e, CEA, CNRS, AIM, 91191 Gif-sur-Yvette, France \label{cea} \and
INAF -- Osservatorio Astronomico di Trieste, via Tiepolo 11, I-34131, Trieste, Italy \label{inaf-trieste} \and
IFPU, Institute for Fundamental Physics of the Universe, Via Beirut 2, 34014 Trieste, Italy \label{ifpu-trieste} \and
Department of Physics; University of Michigan, Ann Arbor, MI 48109, USA \label{michigan} \and
INAF - Osservatorio di Astrofisica e Scienza dello Spazio di Bologna, via Piero Gobetti 93/3, I-40129 Bologna, Italy \label{inaf-oas} \and
INFN, Sezione di Bologna, viale Berti Pichat 6/2, I-40127 Bologna, Italy \label{infn-bo} \and
Dipartimento di Fisica e Astronomia (DIFA), Alma Mater Studiorum - Università di Bologna, via Gobetti 93/2, 40129 Bologna, Italy\label{unibo} \and
Observat\'orio Nacional, Rua General Jos\'e Cristino, 77 – Bairro Imperial de S\~ao Crist\'ov~ao, Rio de Janeiro 20921-400, Brazil \label{obs-brasil} \and
Istituto Nazionale di Astrofisica – Istituto di Radioastronomia (IRA), via Gobetti 101, 40129 Bologna, Italy\label{ira} \and
Dipartimento di Fisica, Sapienza Università di Roma, Piazzale Aldo Moro 5, I-00185 Roma, Italy \label{sapienza} \and
Dipartimento di Fisica “Ettore Pancini”, Università degli studi di Napoli “Federico II”, Via Cintia 21, I-80126 Napoli, Italy\label{uninapoli} \and
Department of Physics, Informatics and Mathematics, University of Modena and Reggio Emilia, 41125 Modena, Italy\label{unimodena} \and
Univ. Lille, Univ. Artois, Univ. Littoral C\^{o}te d'Opale, ULR 7369 -- URePSSS -- Unité de Recherche Pluridisciplinaire Sport Santé Société, F-59000 Lille, France \label{lille} \and
INAF-Osservatorio Astrofisico di Arcetri, Largo Enrico Fermi 5, 50125, Firenze, Italy \label{arceteri} \and
Jodrell Bank Centre for Astrophysics, Department of Physics and Astronomy, The University of Manchester, Manchester M13 9PL, UK\label{manchester} \and
Center for Astrophysics $|$ Harvard \& Smithsonian, 60 Garden Street, Cambridge, MA 02138, USA\label{cfa} \and
HH Wills Physics Laboratory, University of Bristol, Bristol, UK\label{bristol} \and
INFN-Sezione di Genova, Via Dodecaneso 33, 16146, Genova, Italy \label{infn-genova} \and
IRAP, CNRS, Universit\'e de Toulouse, CNES, Toulouse, France \label{irap} \and
California Institute of Technology, 1200 E. California Blvd., Pasadena, CA 91125, USA \label{caltech}}

\abstract
{We investigated the reconstruction of the intra-cluster medium radial density profiles in galaxy clusters, focusing on the impact of cluster morphology, dynamical state, and methodological choices, within the framework of the \chxmate\ project. Using a sample of 109 simulated galaxy clusters analysed along 40 independent lines of sight, we compared azimuthal mean and median techniques for deprojection. We find that azimuthal median-based density profiles recover the true gas density with negligible bias over the radial range $[0.15-1.3] R_{500}$. Mean based profiles are affected by the presence of substructures. While an accurate masking of clumps or substructures can reduce the bias to $< 5\%$, the discrepancy increases substantially in disturbed systems or when substructures remain unresolved.
The mean-to-median density ratio statistically traces projected gas inhomogeneities, although intrinsic scatter and projection effects limit its use for individual clusters. The complementary use of mean and median profiles therefore provides the most complete approach for large X-ray cluster samples.}

\titlerunning{Density profile reconstruction in galaxy clusters}
\authorrunning{I. Bartalucci}

\maketitle
\section{Introduction}\label{sec:intro}

When considering the thermodynamic properties of the hot, ionised intra-cluster medium (ICM), it is typically assumed that the gravitational potential is spherically symmetric and that the gas has settled into hydrostatic equilibrium, where the density and temperature isosurfaces are aligned with the equipotential surfaces. In reality, both X-ray observations and hydrodynamical simulations of galaxy clusters show that the gas and the underlying dark matter distribution are continuously disturbed by the ongoing formation process. 
In cluster cores, departures from equilibrium are often associated with cavities and turbulent gas motions driven by active galactic nucleus feedback (e.g. \citealt{mcnamara2007}; \citealt{fabian2012}; and \citealt{gaspari2014}) or sloshing caused by minor mergers \citep{markevitch2007}. In the outskirts, such departures are mainly linked to ongoing accretion and merger activity (e.g. \citealt{kravtsov2012} and \citealt{walker2019}).
As a result, the ICM is characterised by significant inhomogeneities, in both density and temperature, on a wide range of spatial scales.

Spherical symmetry is commonly assumed when deriving ICM properties from X-ray analyses, as it simplifies the deprojection procedure. It is also expected that geometric biases affecting individual systems average out over large samples because cluster orientations are randomly distributed. The impact of this assumption has been explored in several studies through comparisons of deprojection methods applied to idealised morphologies (e.g. \citealt{binney1978}; \citealt{piffaretti2003}; and \citealt{buote2012a,buote2012b}).  
More recently, the impact of deviations from spherical symmetry on the reconstruction of ICM density profiles has been investigated in work based on cosmological hydrodynamical simulations \citep{2025Veronesi}. These authors show that non-negligible biases in the reconstruction of the mean density profile and its intrinsic scatter occur in the presence of large-scale substructures. These results highlight that projection effects, cluster morphology, and the presence of substructures play a crucial role in shaping the observed density profiles and their statistical properties.

Using numerical simulations, \citet{2013Zhuravleva} showed that for a given radial shell, the density distribution of the ICM can be well described by the combination of a lognormal component, associated with a nearly hydrostatic bulk phase, and a skewed high-density tail produced by inhomogeneities (e.g. \citealt{kawahara2007,vazza2013}). Observationally, gas density profiles are traditionally derived using azimuthally averaged ('mean') surface brightness or emission measure profiles. If not properly accounted for, the presence of inhomogeneities can bias the reconstruction of density profiles, with direct consequences for derived quantities, such as entropy, pressure, gas mass, and importantly, the total mass of the cluster.

The median is expected to trace more closely the peak of the lognormal bulk distribution of the gas density and to not be affected by the high-density tail. Applied to numerical simulations, reconstructed density profiles using the median method have shown excellent agreement with the known \truene, suggesting that this method enables the recovery of the true gas density profile even in the presence of substructures \citep{2015Eckert}. The method has subsequently been applied to observations from ROSAT and \emph{XMM-Newton} \citep[e.g.][]{2015Eckert,2019Ghirardini}.

The level of inhomogeneity in the ICM can be described quantitatively by $C$, the clumping factor. This is defined in simulations as $C = \langle n_{e}^{2}\rangle/\langle n_{e}\rangle^{2}$ \citep{1999Mathiesen}, where the angle brackets ($\langle\rangle$) denote the mean inside a spherical shell. 
\citet{2015Eckert} proposed that the clumping factor could also be recovered observationally from the ratio between the deprojected mean and median density profiles. Clumping factor profiles have been 
derived using this method for several observed clusters up to and beyond $R_{500}$ \citep{2015Eckert,2016Tchernin,2018Ghirardini,2021Mirakhor}.

The clumping factor in the ICM is expected to have two major intrinsic contributors: individual gas clumps and large scale accretion patterns \citep{2013Roncarelli}. However, the imposition of spherical symmetry on a non-spherical object also affects the ratio between the mean and median density profiles.
It is not clear on what scales each contributor dominates, although clumping is likely to be more important at larger radii. Some work has shown the presence of small-scale clumps contributing to approximately half the measured clumping effect at $R_{200}$ \citep{2016Tchernin}. 

Numerical simulations provide a useful framework for disentangling the effect of small- and large-scale inhomogeneities and enable the quantification of the intrinsic limitations of density profile reconstruction when only a single line of sight (LOS) is available. Here, we build upon previous studies and investigate the impact of dynamical state, morphological appearance or parameters, and mass on the reconstruction of ICM density profiles and derived quantities. 
We analyse a sample of simulated galaxy clusters selected to match the mass, redshift, and morphological distributions of the \chxmate\ sample \citep{2021CHEXMATE}. We evaluate different density reconstruction methods and directly compare the reconstructed profiles to the true 3D gas density profiles. We also explore the connection between density reconstruction techniques and gas inhomogeneities.

This paper is organised as follows. In Sect. \ref{sec:300}  we describe the simulated sample and the intrinsic dynamical properties used throughout this work. In Sect. \ref{sec:data_analysis} we present the analysis procedure adopted to extract emission measure profiles and reconstruct density profiles and derived quantities. Section \ref{sec:bias_in_density} studies the impact of different factors on density profile reconstruction, while in Sect. \ref{sec:other_quantities} we discuss the recovery of the logarithmic density gradient and the gas mass. We discuss the relation between the clumpiness and its proxy and the impact on profile reconstruction in Sect. \ref{sec:section_on_clumpiness}, and we draw our conclusions in Sect. \ref{sec:conclusions}. 

Throughout this work we assume a $\Lambda$-Cold Dark Matter ($\Lambda$CDM) cosmological model, with a Hubble parameter $H_{0}$ = 67.7 km s$^{-1}$ Mpc$^{-1}$, a matter density $\Omega_{\rm m}$ = 0.307, and a dark energy density $\Omega_{\Lambda}$ = 0.693. This is also the cosmological model adopted in our simulated sample.

\section{The \simsample\ sample}
\label{sec:300}
\subsection{Sample selection}
\begin{figure}
\centering
\resizebox{0.9\columnwidth}{!}{\includegraphics[]{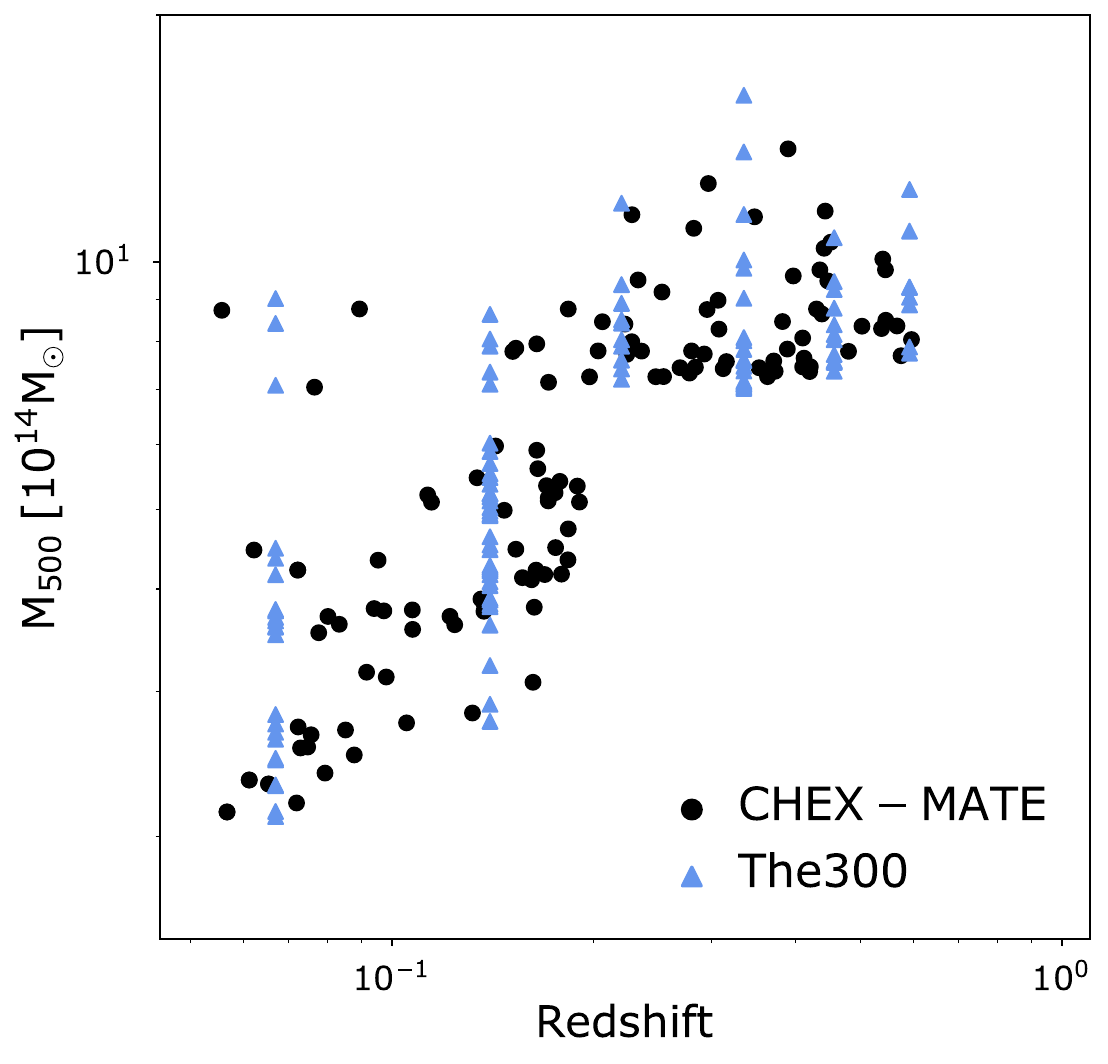}}
\caption{\footnotesize Distribution in the mass-redshift plane of the \chxmate\ sample \citep{2021CHEXMATE} and the matched simulation sample drawn from the \threehun\ suite \citep{2018Cui}, shown with black and blue points, respectively. The masses of \threehun\ clusters are multiplied by 0.8 to facilitate comparison with the \chxmate\ sample (see Sect. \ref{sec:300} for details).}
\label{fig:sample_chex_sim}
\end{figure}

The sample used in this work was drawn from {\sc The Three Hundred} suite of cosmological simulations (\citealt{2018Cui}, \threehunnumbers). These simulations include baryonic processes such as metal-dependent radiative gas cooling, feedback from star formation, supermassive black hole growth, and Active Galactic Nuclei (AGN) feedback \citep{2015Rasia}. The selection criteria used to build this sample are described in detail in Sect.~2 of \cite{2023Bartalucci}. Briefly, the simulated clusters were selected to match the redshift of the \chxmate\ \citep{2021CHEXMATE} sample, composed of 118 clusters selected from the \planck\ PSZ2 catalogue \citep{psz2}, offering a minimally biased view of the cluster population. To match the clusters, in addition to the redshift, we also considered mass and selected simulated clusters with $M_{500}$ close to $M_{\mathrm{YSZ},500}/0.8$, where $M_{\mathrm{YSZ},500}$ is the mass measured from the $Y_{SZ}$ proxy\footnote{The $M_{500}-Y_{SZ}$ relation was calibrated  in \citet{arnaud2010}.} \citep{psz2}. This criterion follows the prescription adopted by the \planck\ collaboration \citep{planck2014}, which assumes a baseline hydrostatic mass bias of 20\% i.e. $(1-b)=0.8$. 
Whenever possible, the simulated clusters were chosen to reproduce the morphological characteristics of the \chxmate\ clusters. The \simsample\ and \chxmate\ clusters are shown in Fig. \ref{fig:sample_chex_sim} with blue and black points, respectively, where the simulated cluster masses are multiplied by a factor of 0.8. The discrete redshift distribution of the \simsample\ sample arises because these objects are GADGET-X resimulations of the MultiDark Planck~2 simulation suite \citep{2016Klypin}, obtained at fixed redshifts: z=[0.07,0.14,0.22,0.33,0.46,0.59].
For each cluster we extracted the following:
\begin{itemize}
\item the radius $R_{500}$\footnote{$R_{\Delta}$ is defined as the radius at which the cluster's total matter density is $\Delta$ times the critical density at the cluster redshift.} and total mass within $M_{500}$;
\item the 3D density profiles, $n_{e,3D}$, extracted in 70 logarithmically spaced radial bins from $0.01R_{500}$ to $7R_{500}$, without excluding accreting substructures;
\item emission-measure maps considering the projection of each cluster along 40 LOSs (as described in Sect.~2 of \citealt{2025Veronesi}; see their Fig.~3 for a visualisation of the LOS construction), with image fields of view spanning $6R_{500}$ and covered by 1200 pixels;
\item the clumpiness profile, extracted in the same radial bins as above, following the prescription given in \cite{2017Planelles}; 
\item dynamical parameters, computed using the 3D information of the simulated particles, as in \citep{2021DeLuca}.
\end{itemize}

The emission measure maps were centred on the peak of the halo total density. This centre, which we refer as the 'theoretical centre', is also the position around which the density and clumpiness profiles were extracted.

\subsection{True 3D density profiles and projected maps}
\label{sec:profsnmaps}
The 3D density profile, \truene, was computed by summing over the mass of all hot gas particles within a given shell divided by the volume of the shell
\begin{equation}
n_{e,\, 3D} = \frac{\Sigma m_{i}}{(4/3)\pi(r_{out}^3-r_{in}^3)},    
\end{equation}
where $r_{in}$ and $r_{out}$ are the inner and outer boundaries of the shell, respectively. Gas was defined as being in the hot phase when its density was lower than the threshold for star formation ($\sim1.95\times10^{-25}$ g cm$^{-3}$), and its temperature above 0.3 keV \citep{2020Ansarifard}. As stated, the pipeline generates profiles starting from $0.01R_{500}$, however, for numerical robustness we considered the minimum radius as that containing at least 100 gas particles. This is typically located at around $0.03R_{500}$.
\threehunnumbers\ sample comprises 115 clusters. Unfortunately, for six clusters the gas particle number is too low in the inner regions and, for this reason, were excluded in this work. Therefore, the sample used for this work comprises 109 clusters in total.

We tested whether there was a relationship between the number of bins containing less than 100 particles in a simulation and a cluster's dynamical and morphological parameters, but no trend was evident. 
We derived the logarithmic slope of the density, \truedlog, and the gas mass profiles, \truemgas, the latter obtained by integrating the \truene\ profiles within concentric shells. We also computed the gas mass enclosed within \Rv, \truemgasv. 

The projected emission measure maps were generated using the Smac code \citep{2005Dolag}, by integrating the contribution of the hot gas particles over a distance of $\pm3R_{500}$ from the cluster centre along 40 LOSs. The conversion from these maps to X-ray surface brightness maps requires the computation of the cooling function, $\Lambda (T,Z)$, which depends on the local temperature and abundance of the gas. \cite{2025Veronesi} shows that, for massive clusters in the soft X-ray band, this function depends weakly, less than $3\%$, on temperature and metallicity. For this reason, we used the emission measure maps directly, without applying a spatially varying conversion.

\subsection{Intrinsic dynamical state parameters}\label{sec:morpho}
The dynamical state of the cluster plays a crucial role in the reconstruction of the density profiles, the deprojection techniques relying on the spherical assumption \citep{2025Veronesi}. 
For this reason, we used a combination of three intrinsic proxies to determine the dynamical state of the \simsaple\ clusters within \Rv. Firstly, we used the mass fraction of all sub-halos in the cluster, $f_{s}$, the sub-halos being identified as structures, with least 20 particles by the Amiga Halo Finder (AHF3) \citep{2009Knollmann}. The $f_{s}$ is defined as
\begin{equation}\label{eq:fs}
    f_{s} = \frac{\Sigma_{i} M_{i}}{M_{\rm 500}},
\end{equation}
where $M_{i}$ is the mass of a sub-halo within \Rv. 
The second dynamical indicator is the scaled offset of the centre of mass with respect to the theoretical centre, $\Delta_{r}$, defined as
\begin{equation}
    \Delta_{r} = \frac{|r_{cm} - r_{c}|}{R_{\rm 500}},
\end{equation}
where $r_{cm}$ is the position of the centre of mass for the cluster, and $r_{c}$ is the theoretical centre of the cluster. 
Finally, we used the virial ratio, $\eta$, defined as
\begin{equation}
    \eta = \frac{2T - E_{s}}{|W|},
\end{equation}
where $T$ is the total kinetic energy, $E_{s}$ is the surface pressure energy from collisionless and gas particles, and $W$ is the total potential energy. For relaxed clusters, $\eta \to 1$.

Since each parameter emphasises a specific aspect in which a cluster can be disturbed \citep{rasia2013}, combining these estimators into a single parameter collects more information and provides a more accurate description of cluster dynamical state \citep{2021DeLuca}. 
We therefore used the relaxation parameter, $\chi$, defined as in \cite{2020Haggar}:

\begin{equation}\label{eq:chi}
    \chi = \sqrt{\frac{3}{\left(\frac{\Delta_{r}}{0.04}\right)^{2}+\left(\frac{f_{s}}{0.1}\right)^{2}+\left(\frac{|1-\eta|}{0.1}\right)^{2}}},
\end{equation}

where the denominators represent typical thresholds to discriminate relaxed from disturbed clusters for each individual parameter. Relaxed clusters minimise the values of $\Delta_r$, $f_s$, and $\eta$ and they are characterised by $\chi > 1$. 

Work on large cluster samples  (e.g. \citealt{2017lovisari, 2019bartalucci,2022Campitiello, ghirardini2022, sanders2025, sanders2026}) have shown that the distribution of morphological parameters is continuous with no obvious bimodal separation between 'relaxed' and 'disturbed' subclasses. For this reason, we defined as the most disturbed (MD) and the most relaxed (MR) clusters the 20 objects with the lowest and highest $\chi$, respectively. Similarly, we defined the clusters most and least affected by substructures as the 20 with the highest and lowest value of $f_s$, respectively. Notice that 20 objects represents $\sim 18\%$ of the sample.

\section{Electron density profile reconstruction from maps}\label{sec:data_analysis}
\subsection{Emission measure profiles}
We extracted the emission measure profiles from the maps described in Sect.~\ref{sec:profsnmaps} following the prescriptions used in \citealt{2023Bartalucci}. 
Briefly, for each object, we identified the peak of the emission within 0.2$R_{500}$ to mimic typical X-ray observational analyses. Note that these centres may not match the centre of the \truene\ profiles described in Sect.~\ref{sec:profsnmaps}.

We then defined concentric annuli using the same radial grid as the \truene\ profiles. The maximum radius of extraction was set to match the average maximum radius, $R_{max} \sim 1.3R_{500}$, probed by \chxmate\ \citep{2023Bartalucci}. 

We determined the emission measure in each annulus using two techniques. Firstly, we computed an azimuthal mean emission measure profile, $\langle EM\rangle$. This is typical for standard X-ray analyses and simply involves summing the intensity of all pixels within a given annulus and dividing this by the number of pixels. From now on, we refer to these profiles as the `mean profiles'. The second method produces an azimuthal median emission measure profile, $\widetilde{EM}$, and is computed by taking the median intensity of all pixels within a given annulus. In cases where a pixel is part of two adjacent bins, we assigned it to the radial bin that contains the pixel centre. This procedure was introduced in order to limit the bias from the emission of substructure too faint to be identified by source detection algorithms \citep{2015Eckert}. We refer to these profiles as the `median profiles'. Quantities derived from mean and median profiles are denoted with the angle brackets ($\langle \rangle$) and tilde ($\; \widetilde{\phantom\ }$), respectively.

\subsection{Substructure masking}\label{sec:masked_maps}
One crucial point of the X-ray analysis based on mean profiles is the identification and excision of clearly visible spatially resolved substructures. \citealt{2025Veronesi} shows that the presence of such obvious substructures can affect the reconstruction of the density profile over the entire radial range considered (see their Fig. B.1). Excision of these regions minimises the bias on the mass derived using the hydrostatic equilibrium assumption. 

Detection tools based on wavelets are routinely used to identify extended sources in X-ray images (e.g. \citealt{finoguenov2020}). However, automatic identification of these structures is difficult due to the abundance of substructure morphologies, their generally low luminosity, and instrumental limitations such as the presence of a high particle background or distortion owing to the point spread function (PSF). For this reason, automatic detections are typically verified by eye in a realistic X-ray analysis.
This procedure would be prohibitive for our sample size, i.e. 40 maps for each of the 109 clusters, but we applied it to a reduced subsample comprised of one randomly chosen projection for each object.
We then extracted the mean profiles using these masked maps. All the results concerning mean profiles in the following are based on these masked profiles unless stated otherwise. For consistency, we extracted the median profiles from the same maps; that is, those used for the mean profiles, but without any mask applied.

The profiles were generated without the addition of any astrophysical background, or any of the instrumental noise. As a result, they are 'ideal observables'. Nevertheless, we note that we compared our density profile reconstruction with those derived from \xmm\ mocks of the \threehun\ clusters (cfr. Sect. 5.1 of \citealt{seppi2026}) and found consistent results.

\subsection{Deprojection technique}

\label{sec:dens_prof_method}

The density profiles were reconstructed from emission measure mean and median profiles using the deprojection with regularisation techniques described in \cite{2006Croston}. The authors tested this method on simulated surface brightness profiles generated from different density profile models, finding good agreement in all cases. They also found excellent agreement with published \emph{Chandra} gas density profiles when testing the method on \emph{XMM-Newton} data of the same clusters. In addition, this method reliably reproduces deprojected \emph{Chandra} density profiles accurately down to an effective radius of $\sim 7$ arcseconds as described in Sect.~4.5 of \citealt{2017Bartalucci}.
As we did not have any convolution by an instrumental spatial response, we applied the \citet{2006Croston} method by performing a simple geometric deprojection of the emission measure profile into 3D assuming spherical symmetry. The density was recovered by taking the square root of the deprojected mean or median emission measure profile. Consistent with the nomenclature, the mean and median density profiles are indicated using $\langle n_{\rm e} \rangle$ and $\widetilde{n_{\rm e}}$, respectively. 

We also used the parametric approach described in \cite{1976Cavaliere} and \cite{2006Vikhlinin} to deproject the emission measure profiles, using the density analytical model based on \citealt{2006Vikhlinin}:
\begin{multline}\label{eq:vikh_model}
    n_e (r) = n_{01} \frac{(r/r_{c1})^{-\alpha}}{(1 + r^2/r^2_{c,1})^3\beta_1/2-\alpha/2} \frac{1}{(1 + r^3/r^3_s)^{\epsilon/3}} + \\ \frac{n_{02}}{(1 + r^2/r^2_{c2})^3\beta_2/3},
\end{multline}
where $n_{01}, r_{c1}, \alpha, \beta_1,r_s,\epsilon, n_{02},r_{c2}$ and $\beta_2$ are free parameters which were fitted via $\chi^2$ minimisation to the emission measure profile. We refer to these density profiles in the following as `parametric density profiles'.

\subsection{Validation of the deprojection method}
\begin{figure}
\centering
\resizebox{0.9\columnwidth}{!}{\includegraphics[]{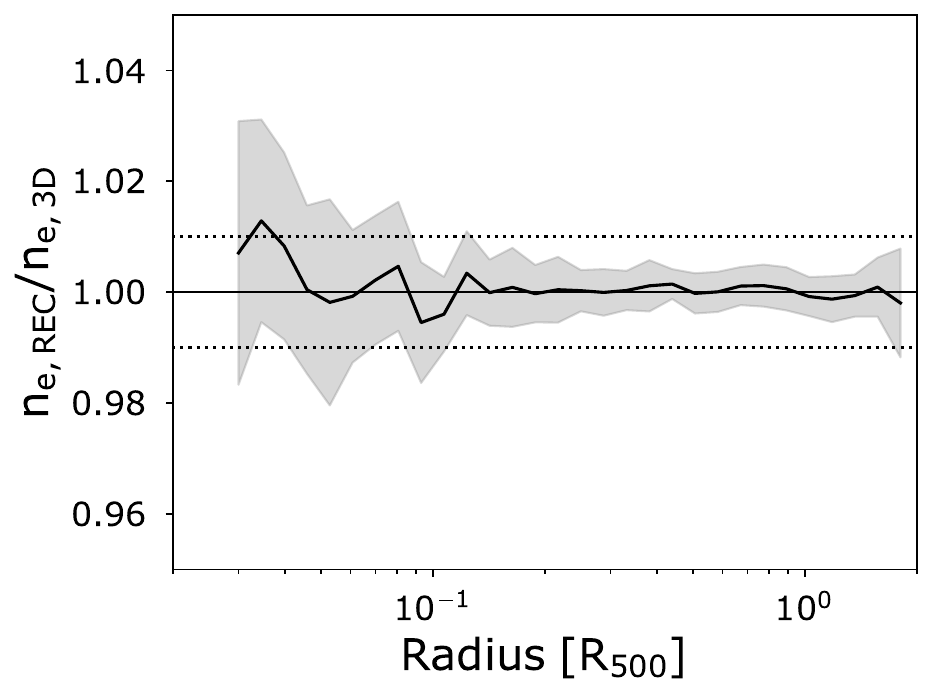}}
\caption{\footnotesize Validation of the deprojection method. The black curve represents the median of the ratios between the recovered density and 3D density profiles. The grey-shaded area represents the median absolute deviation. The dotted lines represent the $+/- 1\%$ levels. }
\label{fig:first_depro_test}
\end{figure}

We first investigated if the non-parametric deprojection could introduce biases by performing a sanity check. We generated a `true' 3D emission measure profile by multiplying the corresponding 3D density profile by the projection matrix of the corresponding 2D density profile. Although the emission measure is a quantity integrated along the LOS, and therefore cannot exist in 3D, the LOS integration is essentially substituted with the sum over all gas particles within the cluster in a series of spherical shells. The integration volume considered for the integration along the LOS is roughly the same, as the integration length for the emission measure images is about the same as the extent of the 3D profiles. We then applied the deprojection algorithm on these reprojected emission profiles and then compared the `recovered' profiles, $n_{\rm e,REC}$, as in Fig. \ref{fig:first_depro_test}, with the \truene. 
The median ratio between the recovered profiles, $n_{\rm e \; 3D, REC}$, and \truene\ profiles and its 68\% median absolute deviation are shown in Fig.~\ref{fig:first_depro_test} with a solid black line and grey-shaded area, respectively. The agreement is excellent within the full radial range, with a maximum deviation of only 1\% at $\sim 0.06$\Rv. 

\begin{figure*}[!ht]
\centering
\includegraphics[scale=0.5]{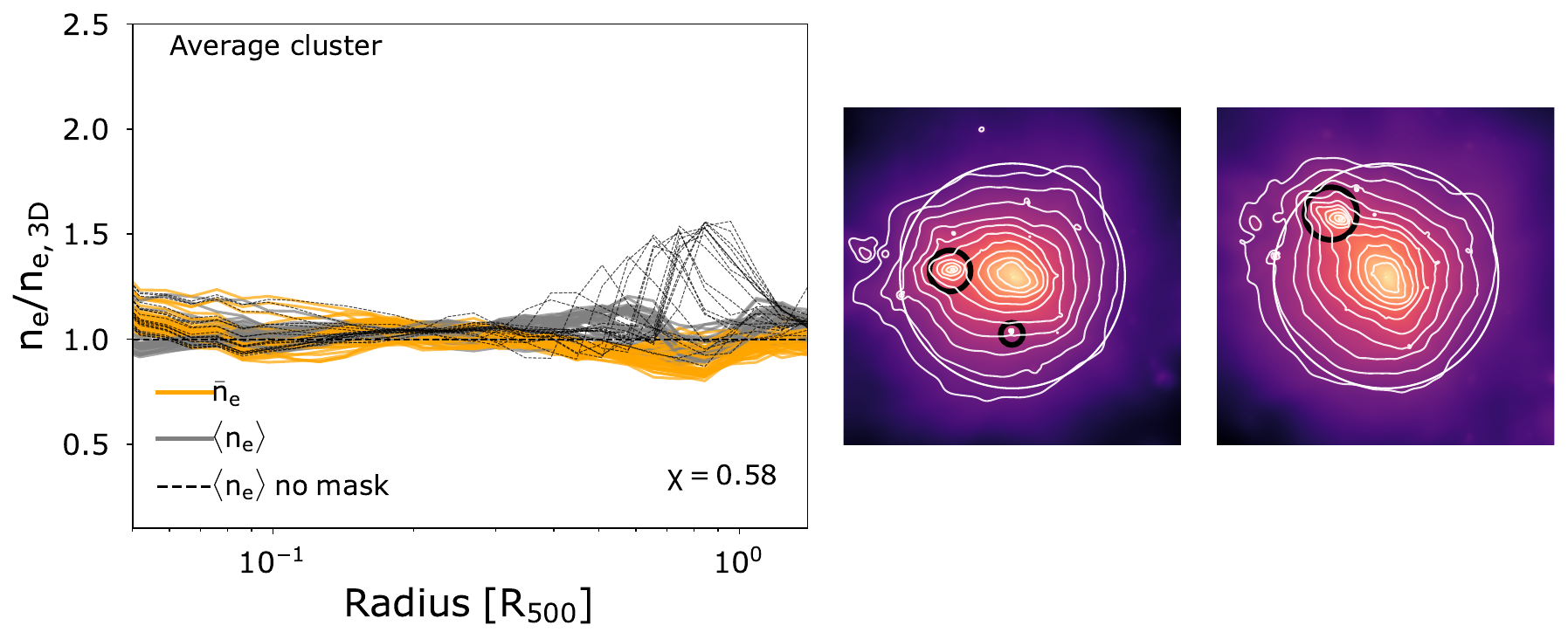}
\includegraphics[scale=0.5]{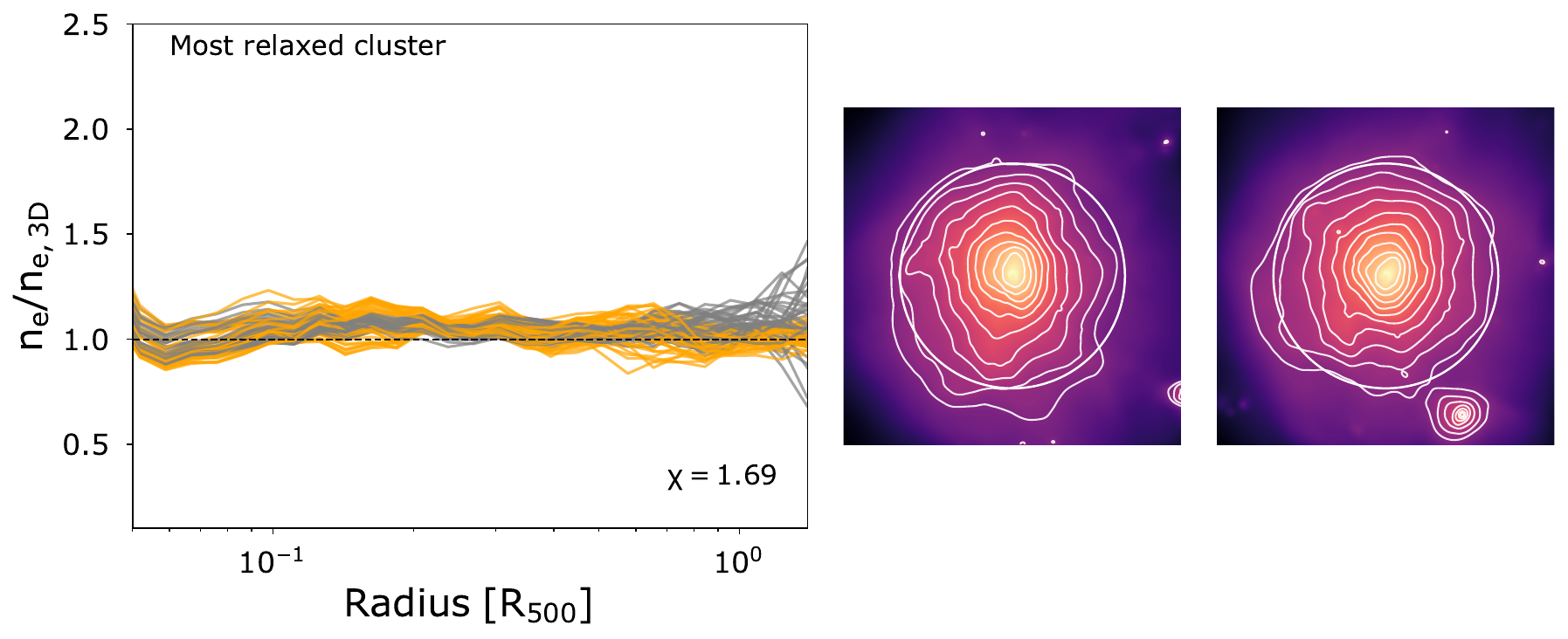}
\includegraphics[scale=0.5]{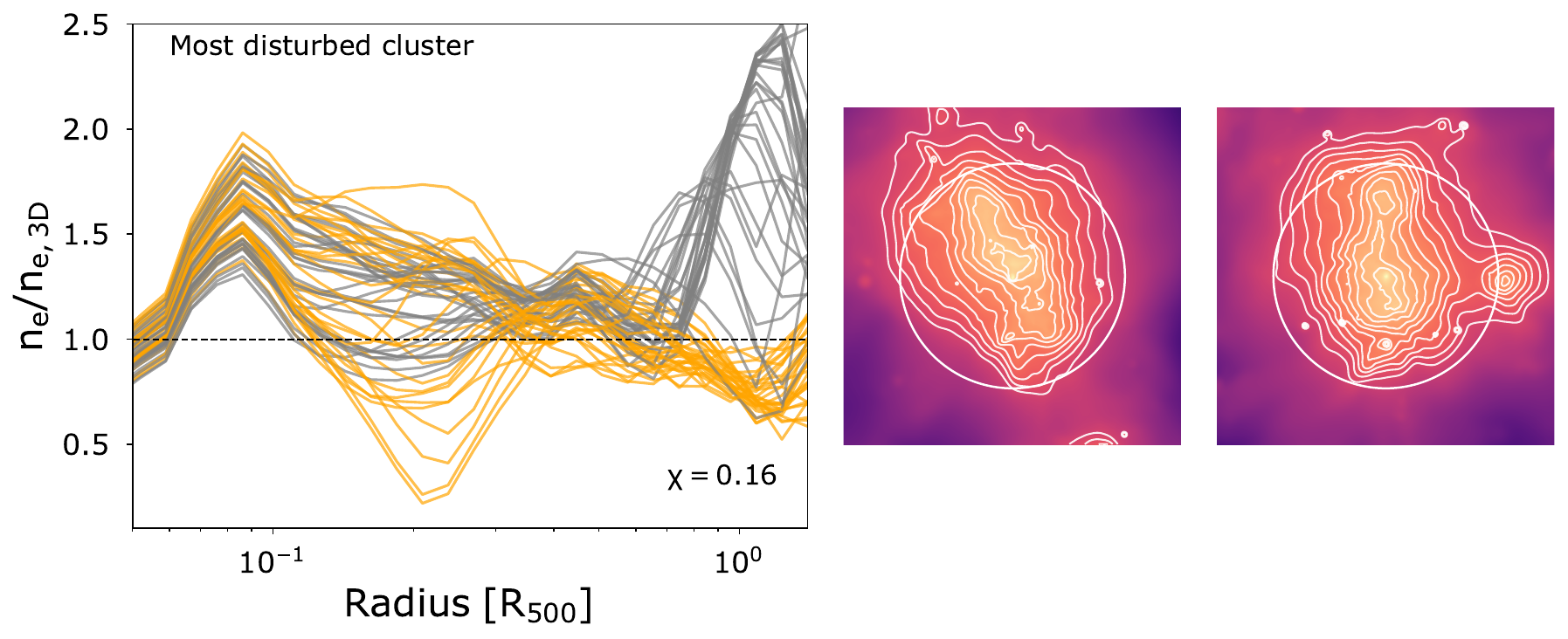}
\caption{\footnotesize Density profile reconstruction of three test case clusters in our sample. The average, MR, and MD clusters are shown in the top, middle, and bottom panels, respectively. For all clusters we also indicate the $\chi$ value, as measured by \cite{2021DeLuca}.  
Top panel, left figure: Ratio between the mean and median and the 3D density profiles for each LOS, shown with solid grey and orange lines, respectively. The dotted black lines in the left figure indicate the ratio between the mean obtained without any masking applied as well as the 3D density profiles. Two right-hand figures: Emission measure maps of the simulated cluster chosen randomly from the 40 LOSs. The white circle indicates $R_{500}$. The masks applied to the substructures are shown with black circles in the right maps. Central and bottom panels: Same as the top panel but illustrating the profiles and the maps of the MR and MD cluster, respectively.}
\label{fig:three_cases}
\end{figure*}

\begin{figure*}[!ht]
\begin{center}
\resizebox{0.95\textwidth}{!}{
\includegraphics[]{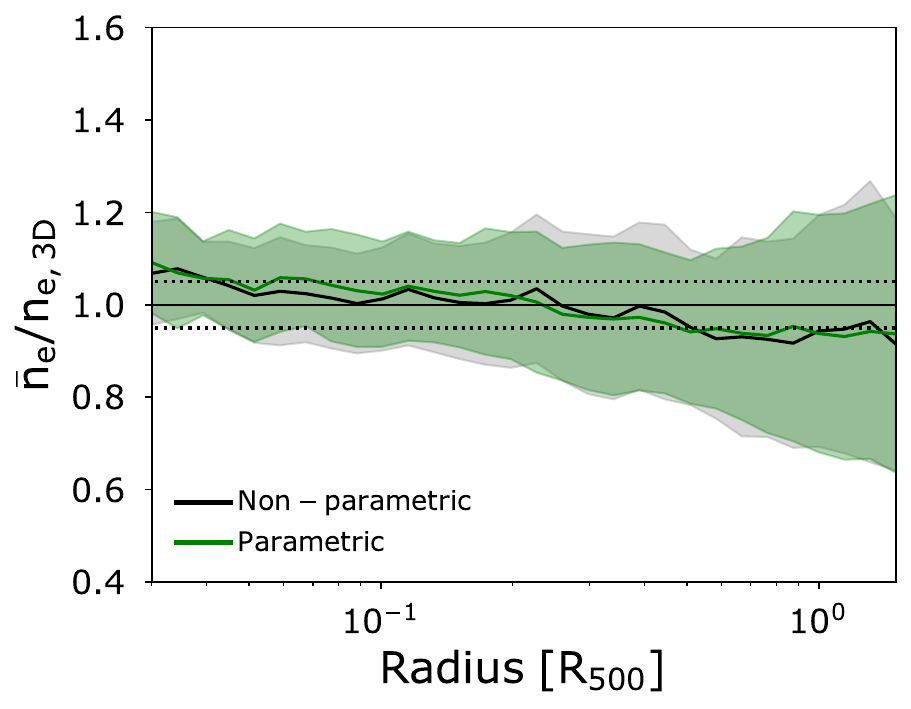}
\includegraphics[]{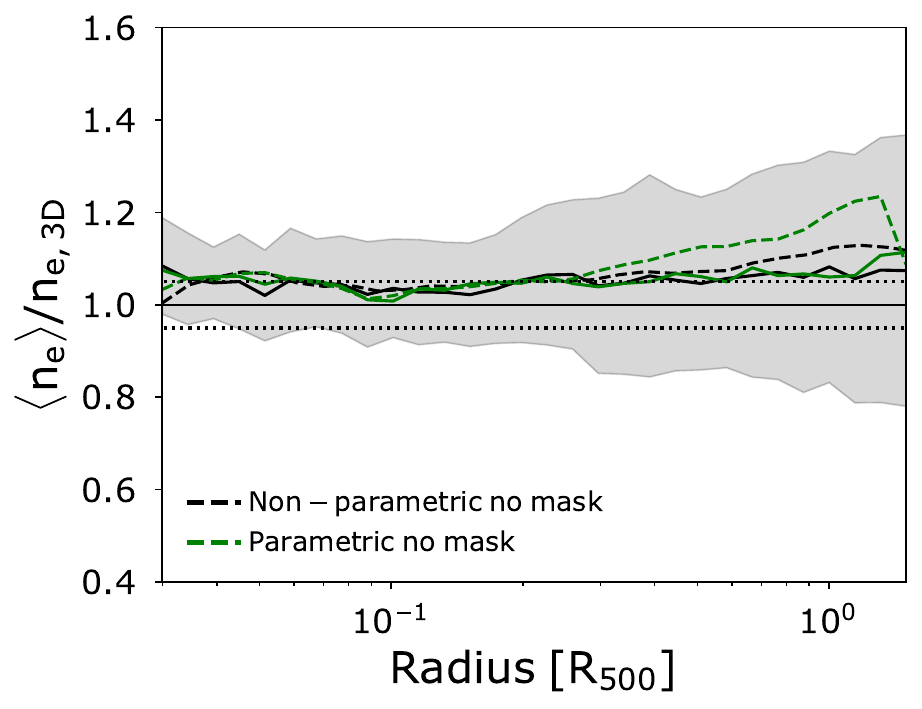}
}
\end{center}
\caption{\footnotesize Biases introduced by the deprojection techniques of the emission measure profiles. Left panel: Median of the ratio between the median non-parametric profiles over the \truene\ considering the full \simsample\ sample (solid black line). The grey-shaded area represents the median absolute deviation. The solid green line and the green-shaded area are the same as the black- and grey-shaded areas, respectively, except that the density profiles were obtained using the parametric technique. The horizontal dotted lines represent the $+/- 5\%$ levels. Right panel: Same as the left panel but illustrating the median ratio of the mean profiles to the \truene\ profiles. The dotted black and green lines are the same as the solid lines, except for the fact that the profiles are extracted from maps where no mask was applied.}
\label{fig:density_mean_median}
\end{figure*}

\section{Sources of bias in density profile reconstruction}\label{sec:bias_in_density}
\subsection{The extreme clusters}
We first investigated the reconstruction of the density profiles for the average and the most extreme objects in our sample in terms of their dynamical state parameters.  
We selected the `average' cluster as that with a $\chi$ value closest to the average $\chi$ of the sample and compared the results to those for the clusters with the highest and lowest $\chi$ of the entire sample, corresponding to the MR and MD objects, respectively. 
Two emission measure maps for each of the three objects were randomly selected from the 40 LOSs and are shown in the right-hand panels in Fig.~\ref{fig:three_cases}. The white circle identifies \Rv.

\subsubsection{The average cluster}\label{sec:most_average}

The top panel of Fig. \ref{fig:three_cases} shows our ability to reconstruct the density profile of the average cluster. The morphology of the object is slightly elliptical, as shown by the two emission measure maps. For this specific system, we excluded the substructures in each of the 40 LOS maps by applying circular masks as represented by the black circles in the figure. 
The ratios between the mean and median density profiles over the \truene\ are shown with solid orange and grey lines, respectively, in the left part of the panel. Only in this case, we also show the ratio between the mean density profiles, obtained applying no mask to the emission measure maps with dotted black lines.
Overall, the median density profiles are in excellent agreement with the \truene, except for a small underestimation of the order of 5\% at $\sim$0.8\Rv. 

The mean density profiles are heavily affected by the presence of the substructure. When they are masked out by deprojection, there is an excellent agreement with the \truene\ except for a $\sim 10\%$ fluctuation at 0.5$\sim$\Rv. This fluctuation may originate from residual emission of the substructure itself. 
The effect of not masking the substructure is clearly visible in the unmasked profile, which shows a 50\% difference from the masked profile at a radius spatially matching the position of the substructure. 

Therefore, the average cluster \truene\ is recovered well by both median and masked mean approaches. Both present fluctuations of at most 5\% and 10\% respectively, with respect to the \truene\ profile. When using the mean profiles, the substructures must be correctly masked before analysis.  

\subsubsection{The most relaxed cluster}\label{sec:most_relaxed}

The middle panel of Fig. \ref{fig:three_cases} shows the results obtained for the MR object of the sample. The morphology of the cluster is characterised by a roundish shape with no evident substructures within \Rv, as shown in the emission measure maps. In this case, both mean and median profiles are in excellent agreement with the \truene. This is not surprising, as the spherical symmetry assumption is an excellent approximation in such a case.

\subsubsection{The most disturbed cluster}\label{sec:most_disturbed}

The bottom panel shows the MD cluster. The maps on the right reveal a complex morphology at all scales, with multiple substructures present both in the core and in the outskirts. 
Substructure masking is not an ideal solution, as it would mask out most of the regions within \Rv. It is worth noting that a typical \chxmate\ X-ray observation does not have sufficient photon statistics or spatial resolution to map such complex ICM morphology of the outskirts. For this reason, this represents the worst-case scenario and highlights how the reconstruction of the density profile is heavily affected by the gas morphology. 

One possible solution for this type of multi-modal object is to employ a model which accounts for the multiple components of the cluster in question. This is beyond the scope of this work, but it will be investigated in future studies employing large simulation datasets.

\subsection{Density reconstruction of the full \simsample\ sample}

The extreme cases highlight how the reconstruction of the density profile strongly depends on the ICM morphology. We now turn our attention to the full sample. 
Unless otherwise stated, all the comparisons presented in this section are based on one randomly selected LOS per cluster. The full set of 40 projections is used in Sect.~\ref{sec:projection_effect} to quantify the scatter induced by projection effects.

\subsubsection{The median density profiles}
We first investigated the reconstruction using the median profiles. For each cluster, we computed the ratio between the reconstructed median density profile obtained from one randomly selected LOS and the \truene. We then estimated the median and the median absolute deviation of these ratios over the full sample of 109 clusters. The resulting ratio profiles are shown in the left panel of Fig.~\ref{fig:density_mean_median}, with black and green lines for the non-parametric and parametric approaches, respectively.

Overall, the agreement is excellent, with a mild bias that spans from $1.03 \pm 0.1$ at $0.05\,R_{500}$ to $0.94 \pm 0.25$ at $R_{500}$, as reported in Table \ref{tab:density_bias_median}. The bias in the centre is linked to the fact that emission measure profiles are centred on the peak of the emission measure. By construction, the deprojected density profile are more peaked in the centre. The centre for profile extraction can also be defined using other quantities, such as the X-ray centroid\footnote{The computation of the emission measure centroid in simulated maps is not representative of the centroid computed on real maps, because of the spatial resolution and presence of multiple substructures. For this reason, we do not use it in this work.} or by leveraging multi-wavelength information such as the position of the brightest cluster galaxy or from the peak of mass from weak lensing mass distribution. A discussion of which of these positions minimises this bias is beyond the scope of this paper. This underestimation at large radii could be related to the large-scale morphology of the ICM and, in particular, to the cluster triaxial shape. In such cases, the azimuthal distribution of the emission is also characterised by large tails, potentially cut out by the median, and the resulting radial profile is slightly underestimated.
 
There is excellent agreement between the results of the parametric and non-parametric profiles, with the values in Table \ref{tab:density_bias_median} being consistent at all radii. This is expected as the computation of the median emission measure tends to smooth out the presence of local inhomogeneities, making the profile more similar to one obtained with a parametric approach. 

\subsubsection{The mean density profiles}\label{sec:mean_density_profile}
The mean profile was extracted from one randomly selected LOS map where substructures were masked. As for the median, we then computed the sample median ratios of the mean density profile to the \truene. These are shown with black and solid green line for the non-parametric and parametric profiles, respectively, in the right panel of Fig.~\ref{fig:density_mean_median}. 
Both parametric and non-parametric profiles yield excellent agreement in the shape of the ratio profile, with a bias of $\lesssim 8\%$ in the [0.03-1.3]\Rv\ radial range (see the radial values in Table \ref{tab:density_bias}). This bias is consistent with that found by, for example, \cite{2025Veronesi} using a different technique to deproject the mean profiles. 

To assess the impact of substructures, we also computed the same quantities but using mean profiles extracted from maps without any mask applied. These are shown with dashed lines. The effect of the substructures is particularly evident at R>\Rv, where the bias increases. Moreover, there is an evident difference between the parametric and non-parametric approaches, the latter presenting a bias of the $20\%$ at R$\sim$1.2\Rv. The effect of a substructure on the reconstruction using a parametric based method is investigated in Appendix~B of \cite{2025Veronesi}, where it is shown that a parametric approach cannot adapt to local inhomogeneities and fits 'bumps' in the radial profile, resulting in an overall overestimation of the density profile. The parametric approach yields similar results to the non-parametric one if the surface brightness or emission measure profile is smooth and free from features such as knees or bumps due to substructures. For this reason, and to more accurately represent the structure we see in the images, without smoothing out features with models, we opted to focus on the non-parametric approach in the rest of the paper.

\subsubsection{Impact of the dynamical state}
\label{sec:dynstate}
\begin{figure}
\centering
\resizebox{0.95\columnwidth}{!}{\includegraphics[]{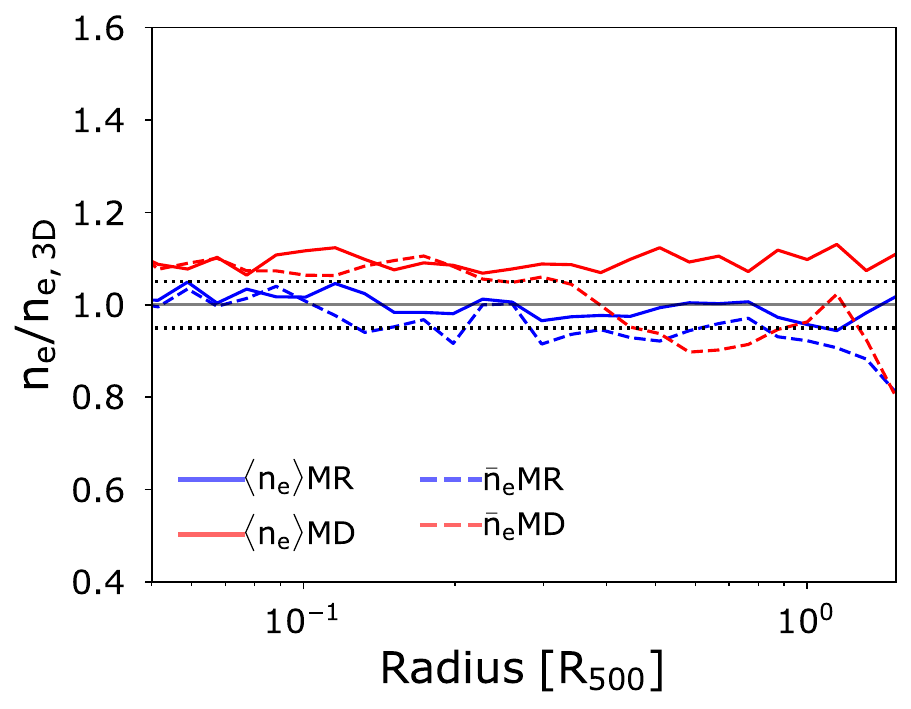}}
\caption{\footnotesize Same quantities shown in the left and right panels of Fig. \ref{fig:density_mean_median}, shown with  dashed and solid lines, respectively. The only difference between the quantities here and those in Fig. \ref{fig:density_mean_median} is that the medians were computed accounting for only the 20 MR and MD objects (shown with blue and red lines, respectively). The horizontal dotted lines represent the $\pm 5\%$ levels.}
\label{fig:morpho_profiles}
\end{figure}
The test cases of the MR and disturbed objects reported in Sects.~\ref{sec:most_relaxed} and \ref{sec:most_disturbed}, respectively, highlight the impact of the dynamical state on the density profile reconstruction. We therefore investigated the reconstruction of the true \truene\ for the 20 MR and MD clusters in our sample. The medians of the ratios between the non-parametric median profiles to the \truene\ are shown in Fig. \ref{fig:morpho_profiles} with blue and dotted red lines for the MR and MD objects, respectively. Overall, for MR objects, the reconstructed median and mean density profiles are in excellent agreement with the \truene, the ratio being close to unity at all radii (see Table~\ref{tab:density_bias_dynamic} for the radial values). For the MD objects, the median profiles present a similar behaviour as for the full sample. The ratio is $1.09 \pm 0.17$ at 0.05\Rv, reducing to $0.96 \pm 0.24$ at \Rv. The mean profiles yield a constant bias of $\sim 1.1$ at all radii. 

\begin{figure*}[!ht]
\begin{center}
\resizebox{0.95\textwidth}{!}{
\includegraphics[]{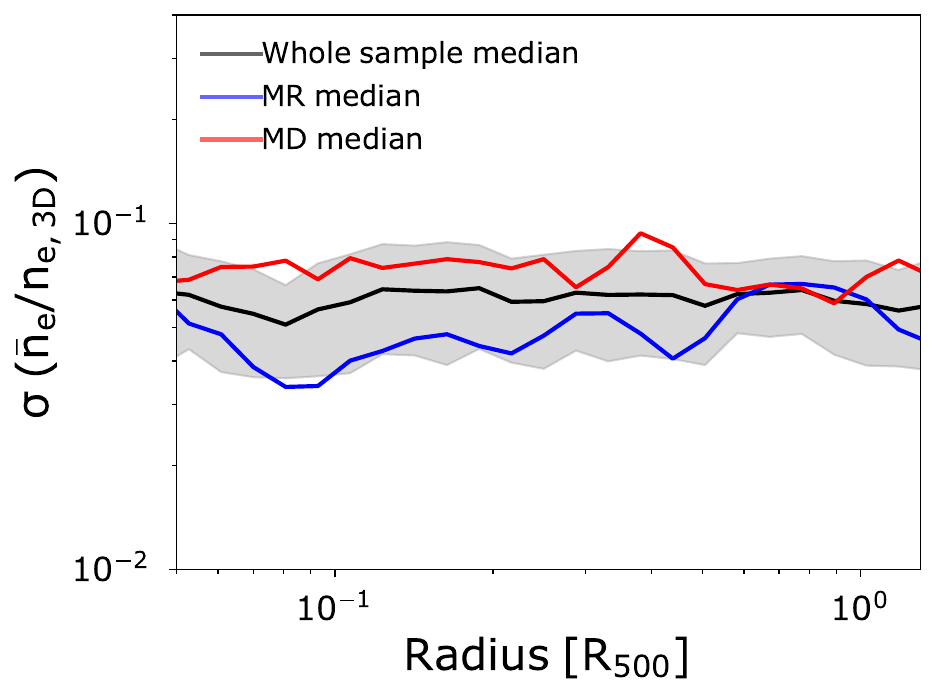}
\includegraphics[]{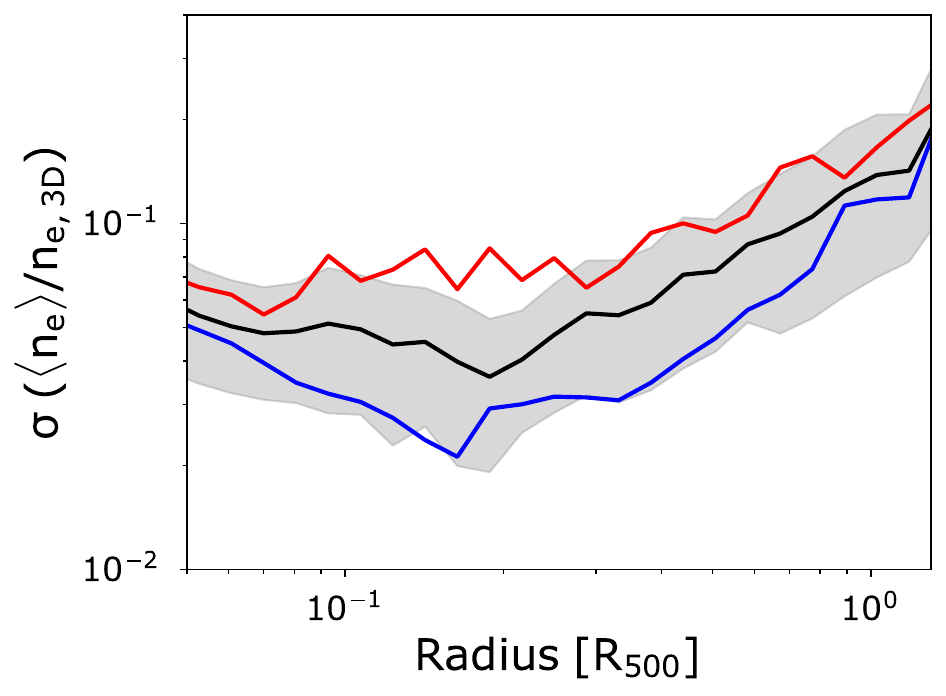}
}
\end{center}
\caption{\footnotesize Scatter between the ratios of the profiles reconstructed from the 40 LOSs over the \truene. Left panel: Intrinsic scatter between the ratio of the median over the 3D density profiles. The red and blue lines represent the median scatter of the MD and MR objects, respectively. The solid black line represents the median of all the profiles shown, and the grey-shaded area represents its median absolute deviation. Right panel: Same as the left panel but illustrating the scatter between the mean density profiles.}
\label{fig:density_scatter}
\end{figure*}

This test shows that the reconstruction bias depends on ICM morphology. For mean profiles, a $\sim 10\%$ bias is observed, which reduces to zero for relaxed objects. The median profiles do not show any differences between relaxed and disturbed objects. The key point for the reconstruction of the profiles is the presence or absence of substructures. As shown for the average cluster in Sect.~\ref{sec:most_average}, the presence of a substructure alone can bias the profile in the annulus containing the substructure by up to $\sim 40\%$.

\subsubsection{The scatter between the lines of sight}\label{sec:projection_effect}

Our sample allows us to investigate the limitation of real observations, which offer only one projection of the cluster. We computed for each cluster the intrinsic scatter between the ratios of the 40 reconstructed LOS profiles to the \truene\ following the method described in \cite{2023Bartalucci}. 

The intrinsic scatters of the median profiles are shown in the left panel of Fig.~\ref{fig:density_scatter}, with the blue and red lines identifying the intrinsic scatter of the samples of the MD and relaxed objects, respectively. Interestingly, the average intrinsic scatter is constant at all radii at $\sim 6\%$, as reported in Table \ref{tab:projection_scatter}. The MD clusters present the largest intrinsic scatter, at $\sim 7\%$. This is due to the fact that the same substructures appear in different position with variable sizes depending on the LOS. For this reason, disturbed clusters with large numbers of substructures are expected to exhibit the largest differences between different LOSs, indicating that they deviate the most from spherical symmetry. 

The same quantities but for the mean profiles are shown in the right panel, using the same colour legend. The results are similar to the median below R=0.5 \Rv, with the scatter range being between $\sigma$ \nemean = 0.05 at $R=0.1\,R_{500}$ and 0.07 at $R=0.5\, R_{500}$. The scatter increases rapidly beyond that radius and is twice as large with $\sigma$ \nemean = 0.13 at $R_{500}$ (cfr. Table \ref{tab:projection_scatter}). This is expected as the mean profiles are more influenced by the presence of substructures triaxial shape and local inhomogeneities. 

These results show that access to only one projection of a cluster limits our capability to reconstruct the underlying density profile by 6\% on average, regardless of the method used, provided that substructures are accounted for. Similar results obtained by \cite{2023Bartalucci} and \cite{2025Veronesi} hints at the same overall small limitation. 

\section{Other quantities}\label{sec:other_quantities}

As stated in the introduction, the density profile is one of the basic ingredients used to characterise the thermodynamic properties of the ICM, including for the calculation of pressure or entropy radial profiles. It is also a key measurement to derive the cluster total mass profile under the assumption of hydrostatic equilibrium or using calibrated proxies. 
In this section we investigate how the density profile reconstruction impacts the derivation of the derivative of the logarithmic gas density profile and gas mass. 

\begin{figure}
\centering
\resizebox{0.95\columnwidth}{!}{\includegraphics[]{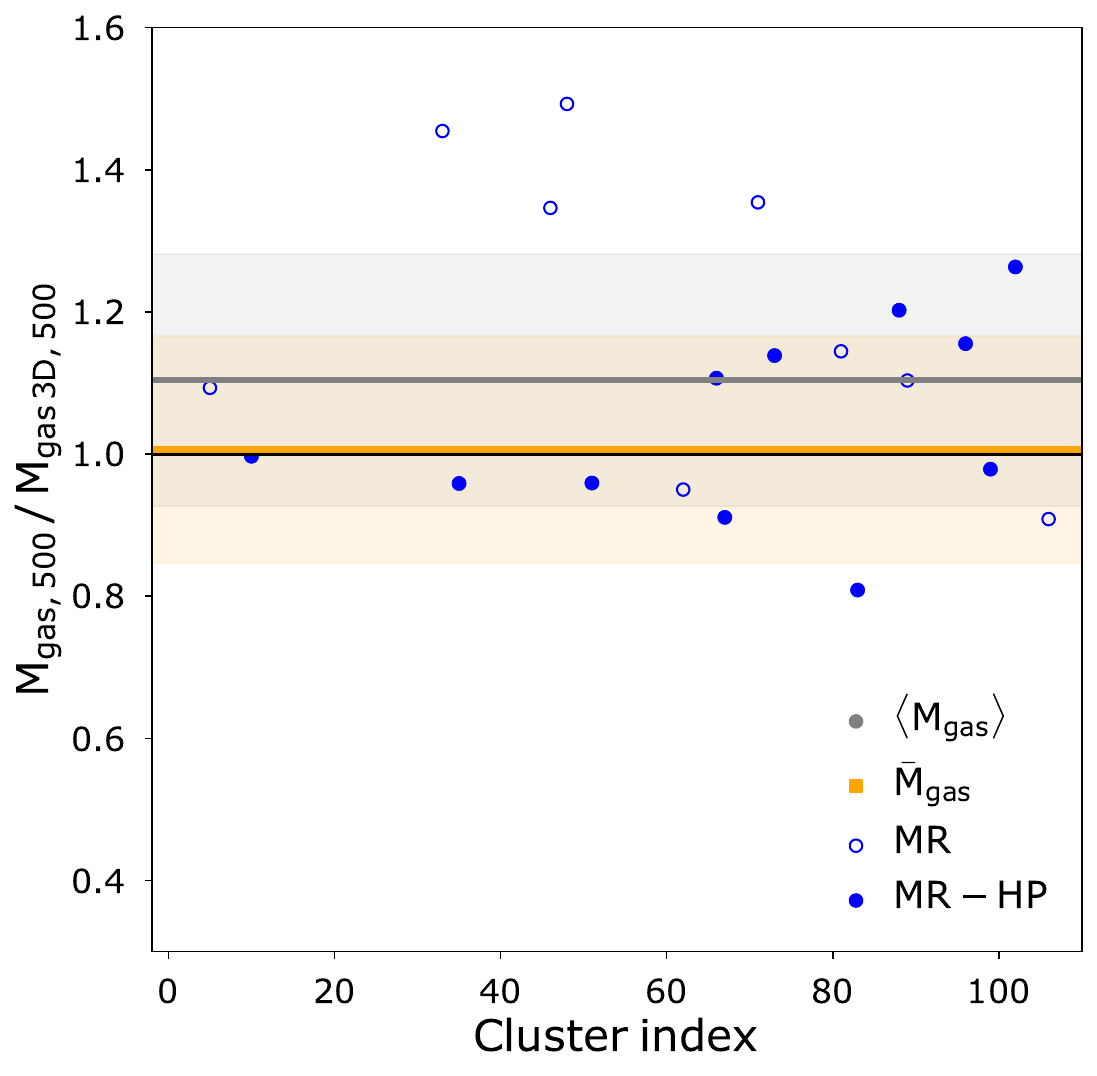}}
\caption{\footnotesize Ratio between the reconstructed gas masses within $R_{500}$ and the true 3D gas masses. The solid yellow and grey lines represent the median ratios obtained using the median and mean density profiles, respectively, with the shaded areas indicating their corresponding median absolute deviations. The open blue circles represent the mass ratios derived from the mean profiles for the MR clusters. The blue-filled circles indicate the subset of the hand-picked (HP) clusters, selected by eye due to their particularly regular morphology, mimicking the selection criteria typically adopted in an X-ray observational analysis.}
\label{fig:gas_mass}
\end{figure}

\subsection{Recovery of the gas mass}\label{sec:gasmass}

The gas mass within \Rv\ represents a key ingredient in cluster cosmology and astrophysics. 
It can be used to derive the proxy $Y_X \equiv M_{\rm gas,500} T_X$, where $T_X$ is the temperature measured within $[0.15-0.75]$\Rv, as introduced by \cite{2006kravtsov}. It can also be used also as an anchor to estimate the non-thermal pressure support using the method described in \cite{2019eckert}. For these reasons, we investigated the impact of the density profile reconstruction technique on the determination of the gas mass within \Rv.

In Fig. \ref{fig:gas_mass}, we present the median of the ratio between the reconstructed gas masses within \Rv\ and the true 3D gas masses ($M_{\rm gas \, 3D,500}$). The horizontal solid yellow and grey lines, along with their corresponding shaded areas, represent the baseline median ratios and median absolute deviations for the entire sample, obtained using the median, $\widetilde{M}_{\rm gas,500}$, and mean, $\langle M_{\rm gas,500} \rangle$, density profiles, respectively. 

The median technique yields a gas mass in excellent agreement with the true value, the median ratio being $(1.01\pm 0.16)$\footnote{The errors reported in this section refer to the mean absolute deviation.}, whereas the gas masses derived from the mean profiles inherits a systematic overestimation, with the median ratio being $(1.10 \pm 0.18)$. The scatter in both cases is similar.

We also investigated the gas mass derived using the mean profiles of the MR cluster subsample discussed in Sect.~\ref{sec:dynstate} only. We remind readers that the \nemean\ profiles were derived masking the substructures. The individual mass ratios derived from the MR mean profiles are plotted as open blue circles in Fig. \ref{fig:gas_mass} and the median ratio and its median absolute deviation (MAD) using only this subsample is $1.11 \pm 0.22$. This behaviour is linked to projection effects along specific LOSs for which even a cluster classified as relaxed present an ellipsoidal morphology. For this reason, we selected among the MR clusters those presenting a roundish morphology similarly to what would be done in a realistic X-ray analysis. These objects, namely MR-hand picked (HP), are shown using the blue-filled circles in Fig. \ref{fig:gas_mass}. 

For this visually filtered sample, the agreement with the $3D$ gas mass is excellent, the median ratio and its MAD being $0.99 \pm 0.16$. This improvement demonstrates that a careful identification of the relaxed systems can mitigate the biases related to \nemean\ profiles\footnote{Note that this was explicitly the approach used to calibrate the $M_{500} - Y_{\rm X}$ relation in for example \citet{2007Arnaud} and \citet{arnaud2010}.}. The numerical results of this section are summarised in Table \ref{tab:gas_mass_stats}.

\subsection{Recovery of the logarithmic slope}\label{sec:gradient}
The derivative of the logarithmic density profile is a key ingredient in the derivation of the cluster total mass under the assumption of hydrostatic equilibrium.
The hydrostatic mass measured from X-ray observations is a local quantity, reconstructed within a particular radius (typically $R_{500}$). It is important that the slope is well constructed around this point.
It is worth noting that the density varies by two orders of magnitude between 0.1\Rv\ and 1\Rv, the typical density ranging from $10^{-2}$cm$^{-3}$ in the core to $10^{-4}$cm$^{-3}$ in the outer parts. 

\begin{figure}
\centering
\resizebox{0.95\columnwidth}{!}{\includegraphics[]{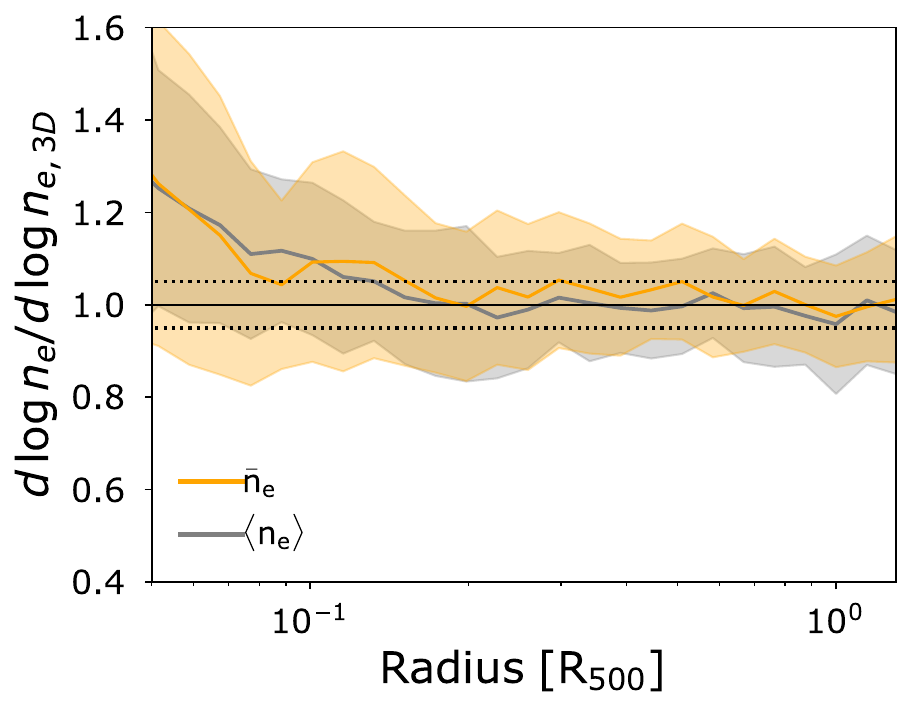}}
\caption{\footnotesize Median of the ratios between the derivative of the logarithmic \nemedian\ (solid orange) and \nemean\ density profiles (solid grey) over the derivative of the logarithmic \truene\ of the whole \simsample\ sample. The orange- and grey-shaded areas represent the median absolute deviation of the \nemedian\ and \nemean, respectively. The horizontal dotted lines represent the $\pm 5\%$ levels. }
\label{fig:gradient_plot}
\end{figure}

Figure~\ref{fig:gradient_plot} reports the medians of the ratio between the recovered logarithmic gradients of the median and mean profiles (orange and black lines, respectively), and the gradients computed using the \truene.
The grey- and orange-shaded area are the median absolute deviations. The behaviour depends on the scale considered. At R>0.2\Rv\ the agreement is excellent both for the mean and median profiles. The mean and median reconstructed profiles are both able to correctly retrieve the shape of the \truene\ in the [0.1-1]\Rv\ radial range. The bias at R<0.2\Rv\ is linked to the X-ray peak used as centre; that is, the gradient is steeper as compared to the true one. This effect affects only the inner regions within which the total mass is generally few percent of the total cluster mass and where the hydrostatic assumption may fail because of the presence of AGN feedback (e.g. \citealt{planelles2014}). 
Interestingly, the medians of the logarithmic gradients of the MR and MD clusters exhibit the same behaviour of the full sample shown in Fig. \ref{fig:gradient_plot}. 

\section{Gas clumpiness and its observational proxy}\label{sec:section_on_clumpiness}
We next quantified the level of ICM inhomogeneity using the clumpiness parameter, $C_{3D}$:
\begin{equation}\label{eq:clumpiness}
  C_{3D} \equiv \frac{\langle \rho^2 \rangle}{\langle \rho \rangle^2} = \frac{\sum_{i}^{N} m_i \rho_i}{\left( \sum_{i}^{N} m_i \right)^2 } V_{shell},
\end{equation}
where $N$ is the total number of the gas particles within the volume of the shell $V_{shell}$ and $m_i$ and $\rho_i$ are the mass and density of the $i^{th}$ fluid element. We also applied the correction for the smoothed-particle hydrodynamics (SPH) computation as in Eq.~6 of \cite{2017Planelles}. We obtained the radial clumpiness profiles, $C$, in the same radial grid used to compute the \truene.

\cite{2015Eckert} proposed that the ratio between density profiles reconstructed from azimuthal mean and median emission measure profiles could be used as an observational proxy, $C_{2D}$, for gas clumpiness:
\begin{equation}
    C_{2D} \equiv \frac{\langle n_e \rangle}{ \widetilde{n}_e }.
\end{equation}
This was proposed under the assumption that the median traces the bulk component of the ICM, while the mean is more sensitive to high-density inhomogeneities.
Here, we investigate the connection between the true 3D clumpiness and its observational proxy, and we assess how both quantities depend on the intrinsic dynamical state and projected morphology of galaxy clusters.

\subsection{Morphological parameters}

The dynamical state of the cluster is reflected to some extent in the morphology of its ICM, although the mapping is not one-to-one due to LOS effects. For this reason, a plethora of morphological parameters are introduced to infer the intrinsic dynamical state of a cluster from its X-ray appearance (e.g. refer to Sect. 2 of \citealt{sanders2025} for a recent catalogue of parameters). Here, we are particularly interested in tracing the large-scale morphology of the ICM, so we focused on two parameters that are among the most sensitive to large scales. First, we selected the ellipticity $\epsilon = a/b$, within \Rv, $a$ and $b$ being the minor and major axis, respectively.  
Then, we considered the mean displacement between the peak and centroid of the emission within N concentric apertures, namely the centroid shift parameter, $\langle w \rangle$ \citep{moh95,bohringer2010}. This is defined as
\begin{equation}\label{eq:centroid_shift}
    \langle w \rangle = \frac{1}{R_\text{500}} \left( \frac{1}{N-1}\sum_{i=1}^N |\underline{R}_i - \underline{R}_\text{peak}|^2 \right)^{1/2},
\end{equation}
where $\underline{R}_i$ is the centroid computed within the $i^{th}$ aperture and $\underline{R}_\text{peak}$ is the peak of the emission and $N$ in our work is set to 10.

\begin{figure*}[!ht]
\begin{center}
\resizebox{1\textwidth}{!}{
\includegraphics[]{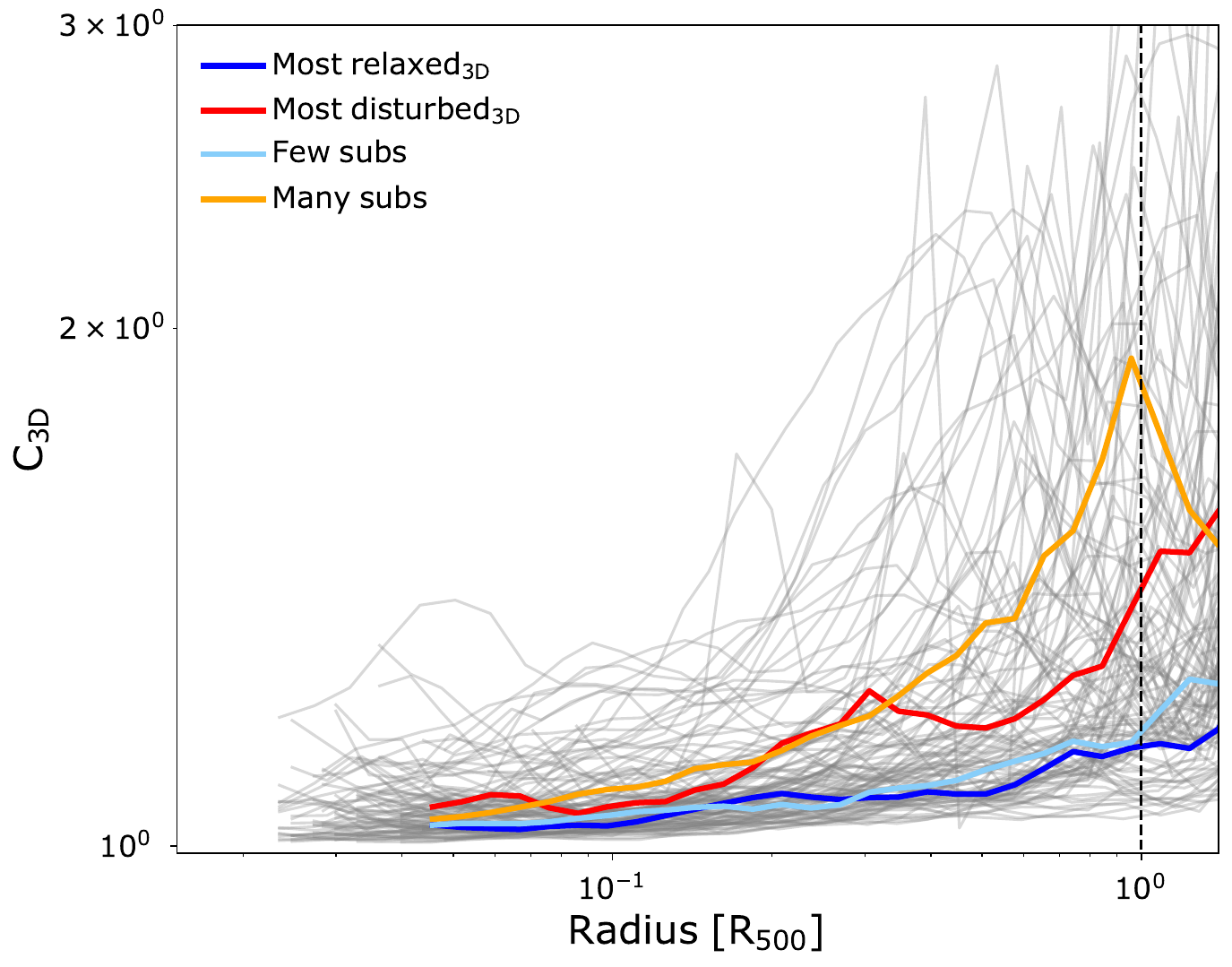}
\includegraphics[]{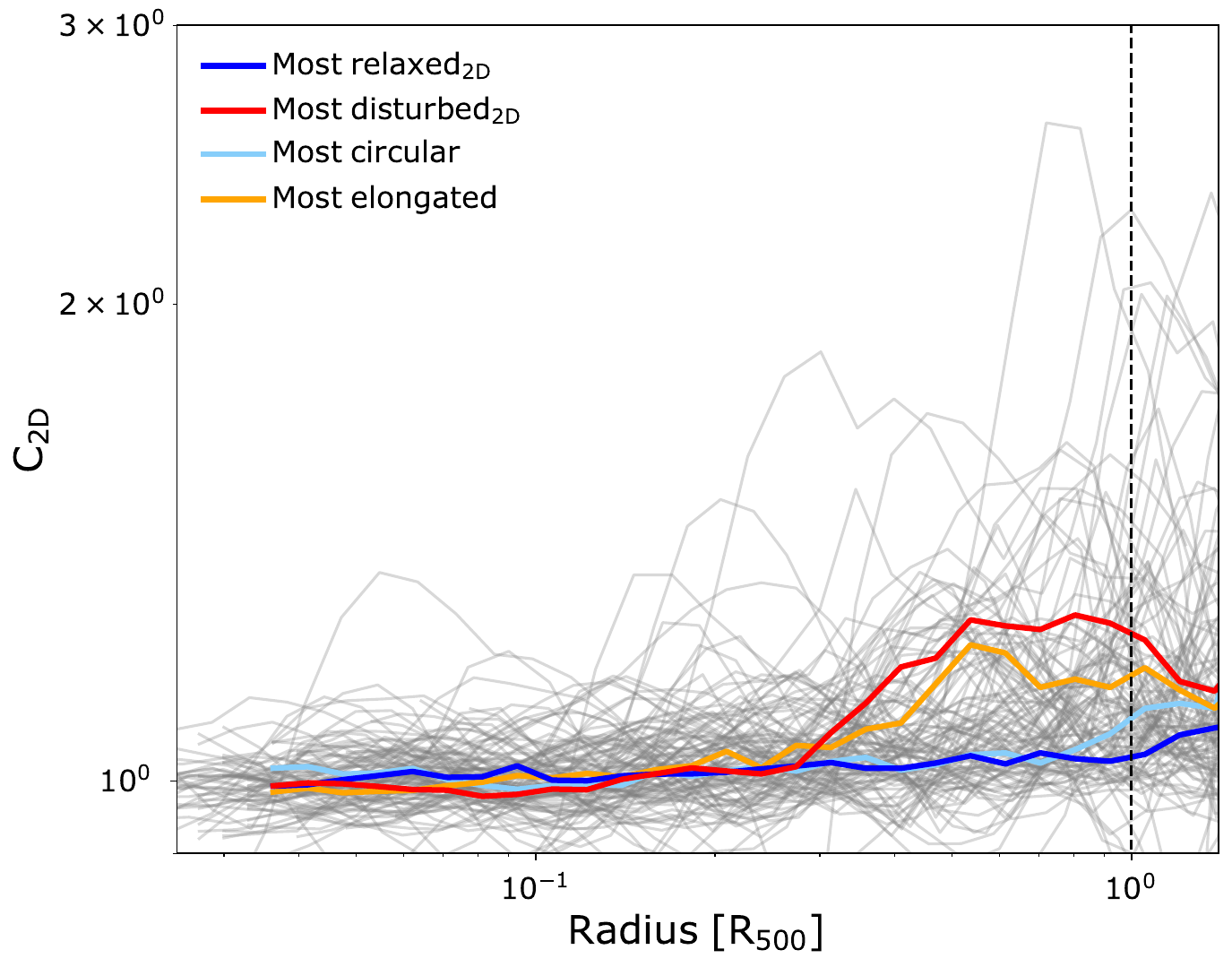}
}
\end{center}
\caption{\footnotesize Behaviour of clumpiness as a function of cluster dynamical and morphological properties. Left panel: Scaled radial profile of the 3D clumpiness measured directly from simulations using Eq. \ref{eq:clumpiness}. The solid blue and red lines represent the median clumpiness profile of the MR and MD clusters, respectively, ranked according to their $\chi$ value. Similarly, the orange and the solid light-blue lines represent the median of the clusters, which contain the largest and smallest number of substructures, respectively, according to the $f_s$ parameter defined in Eq. \ref{eq:fs}. The vertical dashed line marks the position of \Rv. 
Right panel: Radial ratio between the mean and median profiles. The solid red and blue lines correspond to the median of the MD and relaxed clusters, respectively, ranked according to the centroid shift value defined in Eq. \ref{eq:centroid_shift}. The solid orange and light-blue lines represent the median of the most elliptical and most circular clusters ranked according to the ellipticity described in Sect. \ref{sec:morpho}.}
\label{fig:clumpiness_radial_profiles}
\end{figure*} 

We computed these values on the masked maps described in Sect.~\ref{sec:masked_maps} and, similarly to the approach adopted for the 3D dynamical indicators, we selected the 20 most extreme cases. In particular, we defined the objects with the lowest and highest $\langle w \rangle$ as the most 2D relaxed and most 2D disturbed clusters, respectively. Similarly, we defined those with the highest and lowest ellipticity as the most circular and elliptical clusters, respectively. 

\subsection{Radial behaviour and morphology dependence}

The radial behaviour of $C_{3D}$ is shown in the left panel of Fig.~\ref{fig:clumpiness_radial_profiles}. Overall, the clumpiness profiles shown with the solid grey lines increase with radius, in agreement with previous simulation-based studies \citep[e.g.][]{nagai2011,2013Roncarelli,towler2023}. This reflects the transition from the nearly hydrostatic ICM in the cluster cores to the increasingly inhomogeneous gas distribution in the outskirts due to filamentary accretion and infalling substructures.

The separation between the dynamical classes, whose medians are showed with coloured lines, indicates that $C_{3D}$ primarily traces gas density inhomogeneities. The 20 most dynamically disturbed clusters and those hosting the largest fraction of substructures, shown with solid red and orange lines, respectively, exhibit systematically larger $C_{3D}$ than the MR systems and clusters with fewer substructures at all radii. This confirms that the intrinsic clumping factor is mainly driven by the physical processes responsible for generating density fluctuations within the ICM.

The radial behaviour of the observational proxy, $C_{2D}$, is shown in the right panel of Fig.~\ref{fig:clumpiness_radial_profiles}. Despite being derived exclusively from projected quantities, $C_{2D}$ follows a similar qualitative behaviour as the $C_{3D}$. The proxy increases with radius and clearly separates relaxed from disturbed systems, as well as nearly circular from highly elliptical clusters. 

The profiles shown in the left and right panels rely on intrinsic dynamical properties ($\chi$ and $f_s$) measured directly from the simulations and projected morphological estimators ($\langle w \rangle$ and $\epsilon$) that are measured from observations, respectively, however they reveal  similar trends. This shows that the mean-to-median density ratio preserves on average the physical information encoded in the intrinsic clumpiness, while relying exclusively on projected observables.

\subsection{Relation between $C_{3D}$ and $C_{2D}$}
\begin{figure*}[!ht]
\begin{minipage}[t]{0.75\textwidth}
\vspace{0pt}
\resizebox{0.9\textwidth}{!}{
\includegraphics[]{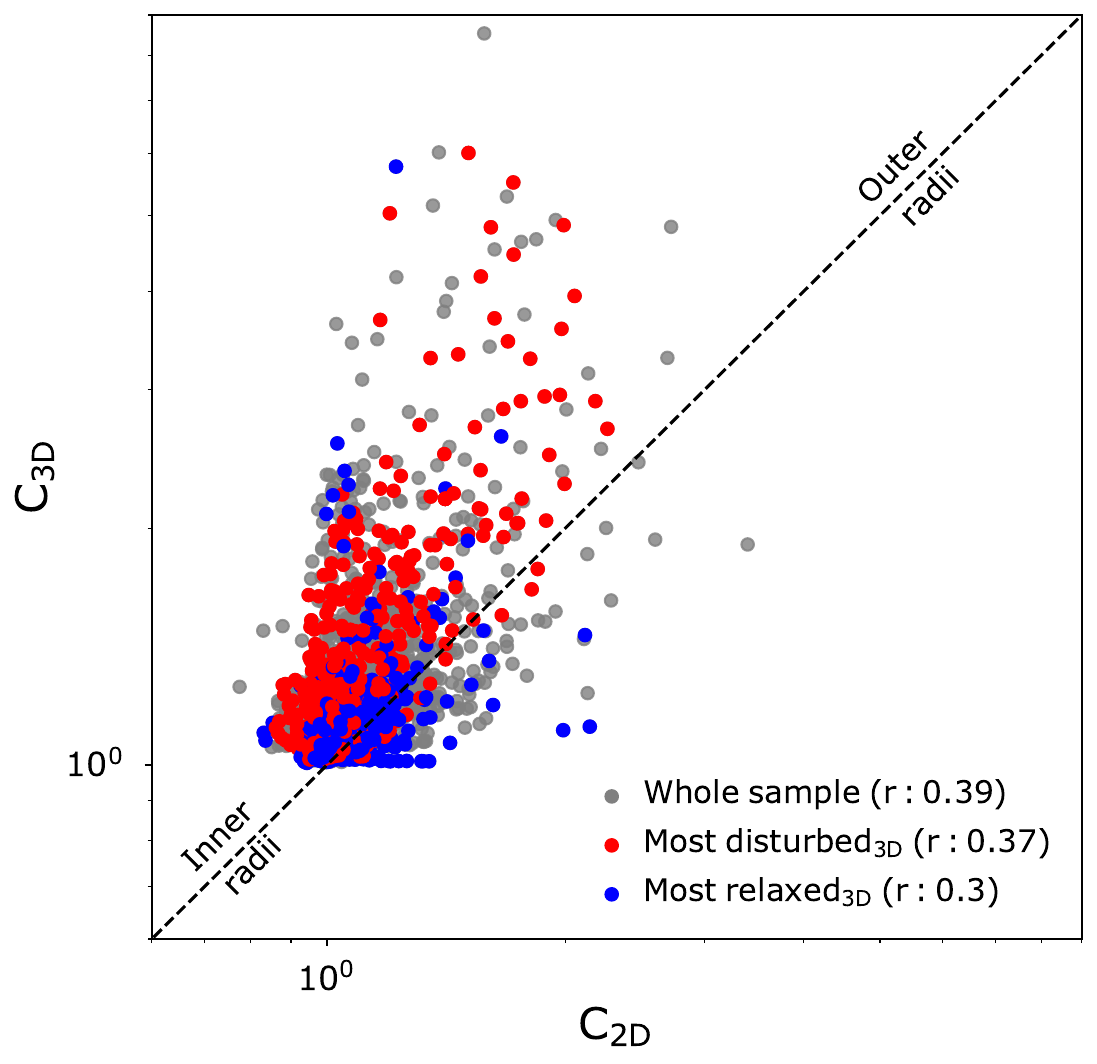}
\includegraphics[]{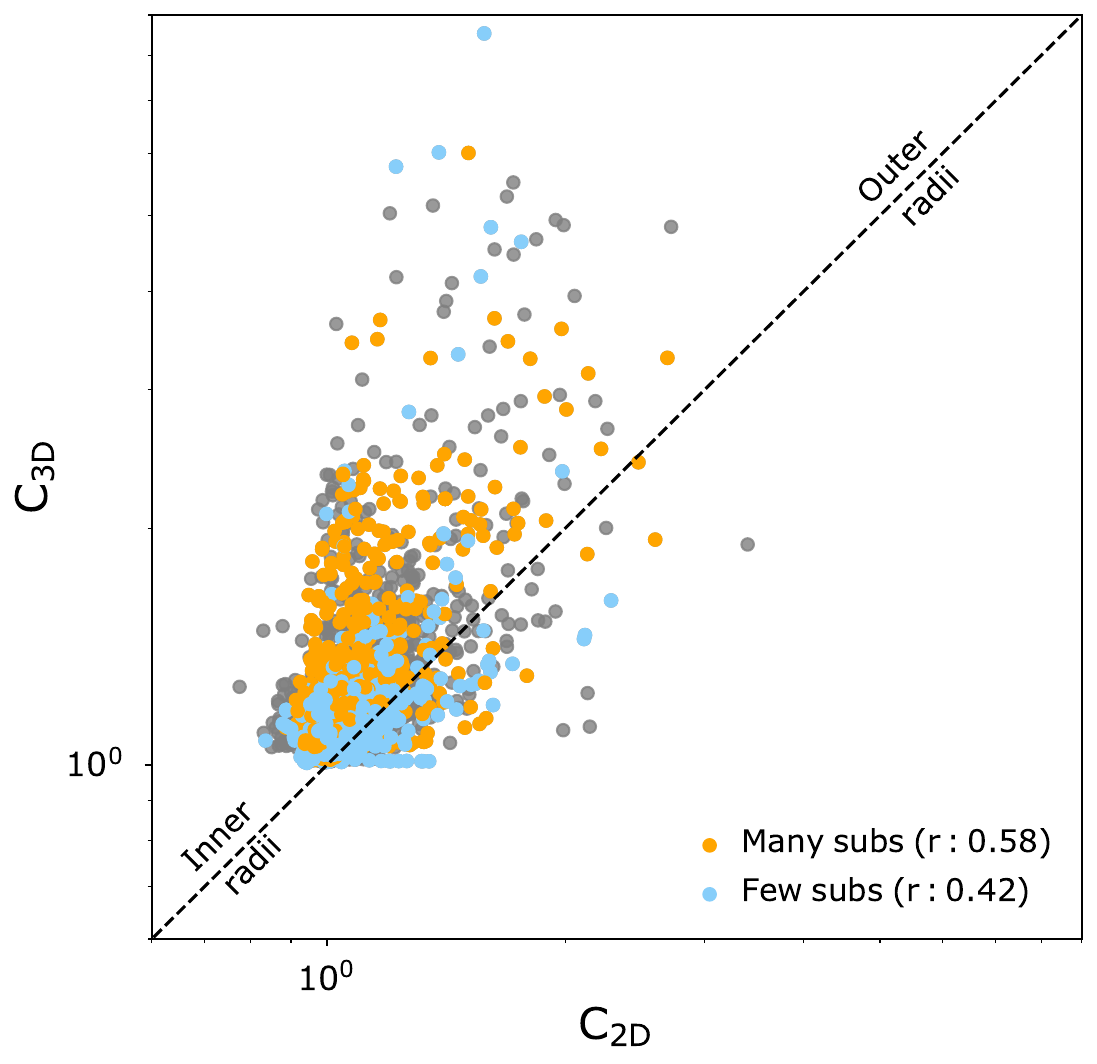}
}
\resizebox{0.9\textwidth}{!}{
\includegraphics[]{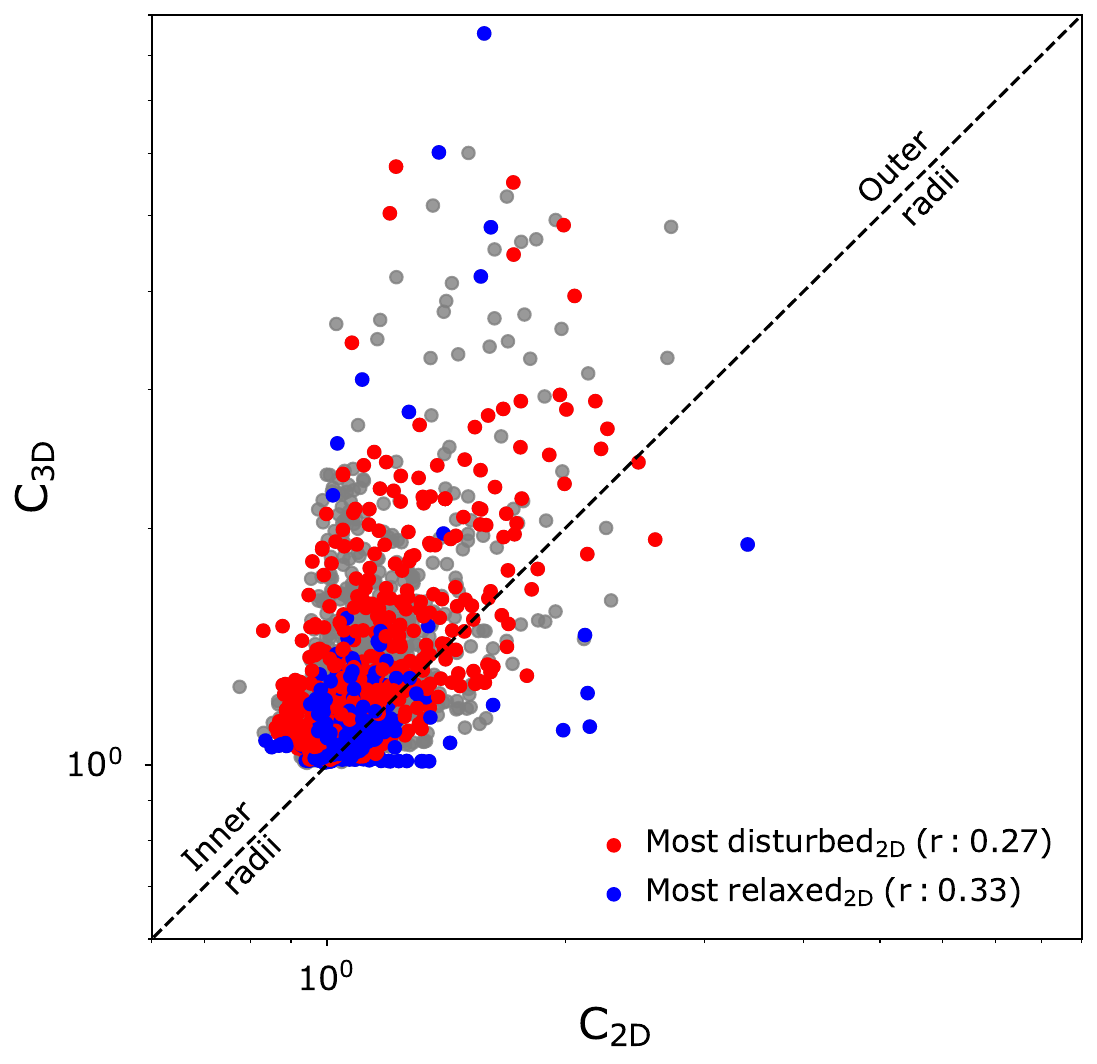}
\includegraphics[]{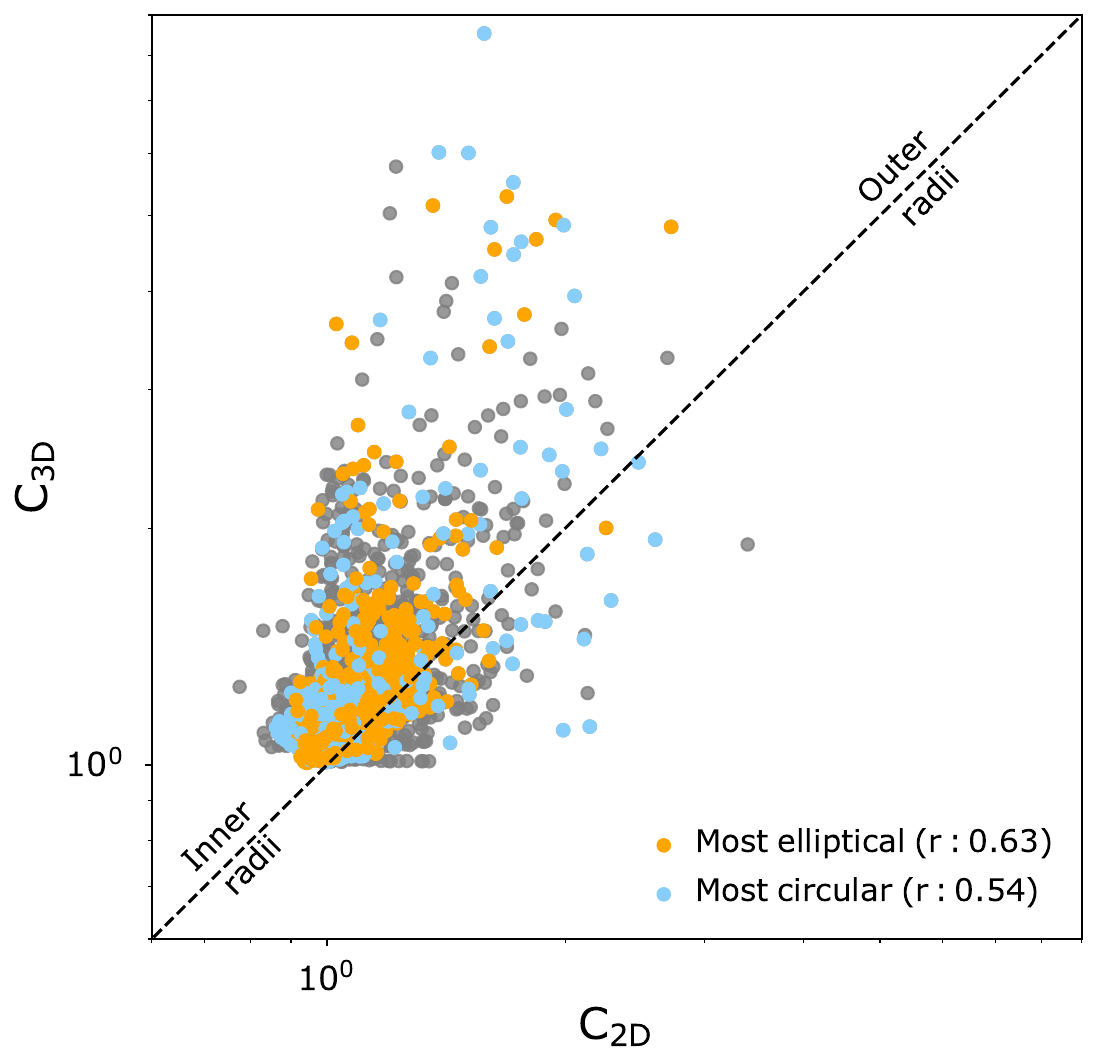}
}
 \end{minipage}\hfill
  \begin{minipage}[t]{0.25\textwidth}
  \vspace{0pt}
 
\caption{\footnotesize 
Correlation between the 3D gas clumpiness and its observational proxy, defined as the ratio between the mean and median density profiles. Each point represents a radial bin in which the ratio was computed. The parameters used to classify the objects are described in Sect. \ref{sec:morpho}. The bottom left part of the plot refers to points extracted from the inner radii of the clusters, while the top right part corresponds to the outer parts. The values in parentheses indicate the Pearson correlation coefficient, r, computed for each subsample. Top left panel: 20 MR (blue points) and 20 MD (red points) objects in 3D according to their $\chi$ parameter. Top right panel: Same as top left but with light-blue and orange points corresponding to the 20 clusters hosting the major and minor number of substructures. Bottom left panel: Same as top left except that the MD and relaxed were defined using the centroid shift parameter $\langle w \rangle$. Bottom right panel: Same as top left but with light-blue and orange points corresponding to the 20 clusters that are the most circular and ellipsoidal according to their ellipticity, $\epsilon$.
}
\label{fig:clumpiness_2d_vs_3d_scatter_plots}
\end{minipage}
\end{figure*}

The correlation between the intrinsic 3D clumping factor and its observational proxy, $C_{2D}$, is shown in Fig.~\ref{fig:clumpiness_2d_vs_3d_scatter_plots}, where the different panels highlight cluster subsamples selected according to their dynamical and morphological properties. A significant scatter is present at all radii, indicating that the observational proxy cannot be used as a quantitative estimator of the intrinsic clumpiness on an object-by-object basis. This scatter arises from projection effects, departures from spherical symmetry, and the limited sensitivity of projected quantities to small-scale density fluctuations \citep{2013Zhuravleva,2016Tchernin}. We therefore investigated whether the relation between $C_{2D}$ and $C_{3D}$ depends on the ICM spatial distribution properties by considering different dynamical and morphological classifications.

To quantify the relation between the $C_{3D}$ and its observational proxy $C_{2D}$ we computed the Pearson correlation coefficient, $r$, for the full sample and for each dynamical and morphological subsample. The complete sample exhibits a moderate correlation ($r=0.39$). Selecting clusters according to their $\chi$ as shown in the top left panel of Fig.~\ref{fig:clumpiness_2d_vs_3d_scatter_plots} does not significantly modify the correlation, with $r=0.37$ and $r=0.30$ for the MD and MR systems, respectively, shown red and blue points. Similarly, the projected dynamical classification based on the centroid shift yields comparable values, with $r=0.27$ and $r=0.33$ for the MD and MR in 2D clusters, respectively, as shown in the bottom left panel.

A stronger correlation is instead found when the sample is divided according to the fraction of substructures or the projected ellipticity as shown in the top right panel. Clusters hosting the largest number of substructures exhibit $r=0.58$, compared to $r=0.42$ for systems containing fewer substructures shown with orange and light-blue points. Likewise, the most elliptical and most circular clusters yield correlation coefficients of $r=0.63$ and $r=0.54$, respectively, shown in the bottom right panel. This suggests that the observational proxy is particularly sensitive to the projected distribution of gas inhomogeneities, which is directly influenced by the presence of substructures and by the large-scale geometry of the ICM.

This behaviour is also reflected in the distribution of the different populations within the $C_{3D}$-$C_{2D}$ plane. The MD clusters, both in terms of the intrinsic dynamical state and the projected centroid shift, together with systems hosting a large number of substructures, preferentially populate the region of high $C_{3D}$ and high $C_{2D}$, while relaxed systems are generally associated with lower values of both quantities. Conversely, the two ellipticity-selected populations largely overlap despite exhibiting the strongest correlations. This indicates that the projected shape alone is not sufficient to uniquely determine the intrinsic clumpiness, although it contributes significantly to the value of the observational proxy.

Overall, these results indicate that $C_{2D}$ should not be interpreted as a direct measurement of $C_{3D}$. It does provide a statistical description of projected gas inhomogeneities, retaining sensitivity to the average level of gas clumping across cluster populations while being significantly affected by projection effects on individual objects.

\section{Summary and conclusions}\label{sec:conclusions}

In this work we investigated the factors influencing the reconstruction of the density profiles by analysing a sample of 109 simulated galaxy clusters whose mass, redshift, and morphological distributions were matched to those of the \chxmate\ sample. For each cluster, emission measure maps were extracted along 40 independent LOSs, allowing us to quantify the impact of projection effects, cluster morphology, and methodological choices on density reconstruction and on derived quantities.

Our main findings can be summarised as follows:

\begin{itemize}

\item The parametric and non-parametric deprojection techniques provide consistent reconstructions of the ICM density profiles when using the azimuthal median profiles \nemedian. They are consistent with the input value within the median absolute devitation (see Table \ref{tab:density_bias_median}) over the $[0.05-1.2]R_{500}$ radial range. However, the parametric reconstruction method becomes increasingly biased when using the mean profiles \nemean\ in cases where substructures are inadequately masked or in the presence of local inhomogeneities.

\item The profiles reconstructed using the median method remain consistent with the true density at all radii, with median ratios between $1.01\pm0.11$ at $0.1R_{500}$ and $0.94\pm0.25$ at $R_{500}$. Mean profiles reconstructed after masking substructures are also consistent with the \truene\ with a small overestimation of $\sim5$-$8\%$, with ratios ranging from $1.04\pm0.11$ to $1.08\pm0.25$ over the same radial range (Tables \ref{tab:density_bias_median} and \ref{tab:density_bias}). The dominant sources of systematic uncertainty are cluster morphology and unresolved substructures rather than the deprojection algorithm itself. Disturbed systems exhibit systematically larger biases than relaxed clusters, while the absence of substructure masking increases the mean-profile bias from $\sim6\%$ to $\sim13\%$ at $R_{500}$ and up to $23\%$ at $1.2R_{500}$ (cfr. Table \ref{tab:density_bias_dynamic}).

\item Projection effects introduce only a modest intrinsic uncertainty in the reconstructed density profiles. The median intrinsic scatter remains nearly constant at $\sigma_{\tilde n_e}=0.06\pm0.02$ over the full radial range, whereas the mean profiles become progressively more sensitive to projection effects, increasing from $\sigma_{\langle n_e\rangle}=0.05\pm0.02$ at $0.1R_{500}$ to $0.15\pm0.07$ at $1.2R_{500}$ (Table \ref{tab:projection_scatter}).

\item The gas mass within $R_{500}$ inherits the systematic behaviour of the underlying density profiles. The median profiles recover the true gas mass with essentially no bias ($M_{\rm gas}/M_{\rm gas,3D}=1.01\pm0.16$), whereas the mean profiles overestimate the true gas mass by $10\%$ ($1.10\pm0.18$). However, the gas mass derived using the \nemean\  becomes fully consistent with the true value, the median ratio being ($0.99\pm0.16$; Table \ref{tab:gas_mass_stats}), if a careful selection of the MR objects is made.

\item The logarithmic density gradient is accurately recovered outside the central region by both \nemean\ and \nemedian. Beyond $0.2R_{500}$, the reconstructed slopes remain within approximately $5\%$ of the true values, with ratios of $1.02\pm0.16$ ($\tilde n_e$) and $0.99\pm0.13$ ($\langle n_e\rangle$) at $0.25R_{500}$, and remaining consistent with unity out to $1.2R_{500}$ (Table \ref{tab:slope_accuracy}). Therefore, both methods are suitable for hydrostatic mass reconstruction outside the cluster core.

\item The ratio between the mean and median density profiles, $C_{2D}$, cannot be interpreted as a direct estimator of the intrinsic 3D clumping factor, $C_{3D}$, on an object-by-object basis. However, it provides a statistical tracer of projected gas inhomogeneities, remaining sensitive to cluster populations while being affected by projection, substructures, and departures from spherical symmetry.

\end{itemize}

Overall, our results demonstrate that azimuthal median profiles provide a robust tool for reconstructing the bulk ICM density over a wide radial range, while mean profiles remain essential for probing the innermost regions of galaxy clusters, where the median technique is limited by spatial resolution. For relaxed systems, which present a regular morphology, the mean and median profiles yield identically unbiased estimates of the gas mass. Both methods can be used as input to obtain the total mass profile under the hydrostatic assumption, yielding unbiased results in both cases. We therefore advocate the complementary use of both approaches, depending on the scientific goal. These findings provide quantitative guidance for current and future X-ray studies of galaxy clusters, including large observational programmes such as \chxmate.

\begin{acknowledgements}
This work has been made possible by The Three Hundred (https://the300-project.org) collaboration. The simulations used in this paper have been performed in the MareNostrum Supercomputer at the Barcelona Supercomputing Center, thanks to CPU time granted by the Red Española de Supercomputación. As part of The Three Hundred project, this work has received financial support from the European Union’s Horizon 2020 Research and Innovation programme under the Marie Sklodowskaw-Curie grant agreement number 734374, the LACEGAL project. ER is supported by NASA EGIP Grant 80NSSC25K0006. This research was supported by the International Space Science Institute (ISSI) in Bern, through ISSI International Team project \#565 ({\it Multi-Wavelength Studies of the Culmination of Structure Formation in the Universe}). AB acknowledges support from the ERC CoG $\vec B$ELOVED, GA N.101169773. MS acknowledges financial contributions from contract INAF mainstream project 1.05.01.86.10, INAF Theory Grant 2023: Gravitational lensing detection of matter distribution at galaxy cluster boundaries and beyond (1.05.23.06.17), INAF Guest Observer Grant 2024: Towards anchoring the mass scale of galaxy clusters with galaxy kinematics (1.05.24.02.15), and contract Prin-MUR 2022 supported by Next Generation EU (n.20227RNLY3, The concordance cosmological model: stress-tests with galaxy clusters). GWP, MA, RTD, EB, EP, and ML acknowledge support from the Centre national d’études spatiales (CNES), France (ROR: https://ror.org/04h1h0y33) within the framework of the XMM-{\it Newton} mission.
JS was supported by NASA grants  80NSSC25K0006 and 80NSSC25K8009.
MG acknowledges support from the ERC Consolidator Grant BlackHoleWeather (101086804).
HB, FDL, and PM acknowledge the financial contribution from the contract Prin-MUR 2022 supported by Next Generation EU (M4.C2.1.1, n.20227RNLY3 \textit{The concordance cosmological model: stress-tests with galaxy clusters}) and the support by INFN through the InDark initiative.
RMB acknowledges financial support from the National Council for Scientific and Technological Development (CNPq) Grant No. 447475/2024-0.
BJM acknowledges support from Science and Technology
Facilities Council grants ST/V000454/1 and ST/Y002008/1.

\end{acknowledgements}

\bibliography{bib}
\clearpage
\begin{appendix}

\section{Implications for observations}\label{sec:appendix_prespr}
In this Appendix we summarise results obtained in this work in terms of systematic effects that can be associated with density profile reconstruction and the derived quantities as respect to the \truene.
\subsection{Systematic biases in density profiles}
Table~\ref{tab:density_bias} summarises the systematic bias on reconstructed density profiles interpolated at fixed radii based on the results shown in Fig. \ref{fig:density_mean_median}.

\begin{table}[h]
\caption{Bias of median density profiles.}
\label{tab:density_bias_median}
\centering
\resizebox{0.7\columnwidth}{!}{%
\begin{tabular}{lcc}
\hline\hline
& \multicolumn{2}{c}{\nemedian/\truene} \\
\cline{2-3}
Radius &
Non-parametric &
Parametric \\
\hline
$0.05\,R_{500}$ & $1.03 \pm 0.10$  & $1.04 \pm 0.11$ \\
$0.1\,R_{500}$ & $1.01 \pm 0.11$ & $1.02 \pm 0.11$ \\
$0.25\,R_{500}$ & $1.01 \pm 0.16$ & $0.99 \pm 0.15$ \\
$0.5\,R_{500}$ & $0.96 \pm 0.17$ & $0.94 \pm 0.15$ \\
$0.75\,R_{500}$ & $0.93 \pm 0.21$ & $0.93 \pm 0.21$ \\
$1.0\,R_{500}$ & $0.94 \pm 0.25$ & $0.94 \pm 0.26$ \\
$1.2\,R_{500}$ & $0.95 \pm 0.28$ & $0.94 \pm 0.27$ \\
\hline
\end{tabular}%
}
\vspace{1mm}
\tablefoot{The listed values are the median and median absolute deviation of the ratio between the reconstructed median density, \nemedian, and the true 3D density, \truene, at fixed radii for the non-parametric and parametric deprojection methods.}
\end{table}

\begin{table}[h]
\caption{Bias of mean density profiles.}
\label{tab:density_bias}
\centering
\resizebox{\columnwidth}{!}{%
\begin{tabular}{lcccc}
\hline\hline
& \multicolumn{4}{c}{\nemean/\truene} \\
\cline{2-5}
Radius &
Masked &
Not masked &
Parametric masked &
Parametric not masked \\
\hline
$0.05\,R_{500}$ & $1.03 \pm 0.10$ & $1.07 \pm 0.09$ & $1.04 \pm 0.12$ & $1.07 \pm 0.09$ \\
$0.1\,R_{500}$ & $1.04 \pm 0.11$ & $1.03 \pm 0.03$ & $1.01 \pm 0.11$ & $1.02 \pm 0.05$ \\
$0.25\,R_{500}$ & $1.07 \pm 0.23$ & $1.05 \pm 0.03$ & $1.05 \pm 0.16$ & $1.05 \pm 0.05$ \\
$0.5\,R_{500}$ & $1.05 \pm 0.19$ & $1.07 \pm 0.04$ & $1.06 \pm 0.19$ & $1.12 \pm 0.05$ \\
$0.75\,R_{500}$ & $1.07 \pm 0.23$ & $1.11 \pm 0.06$ & $1.07 \pm 0.22$ & $1.14 \pm 0.07$ \\
$1.0\,R_{500}$ & $1.08 \pm 0.25$ & $1.13 \pm 0.08$ & $1.06 \pm 0.22$ & $1.20 \pm 0.11$ \\
$1.2\,R_{500}$ & $1.06 \pm 0.27$ & $1.12 \pm 0.08$ & $1.08 \pm 0.24$ & $1.23 \pm 0.12$ \\
\hline
\end{tabular}%
}
\vspace{1mm}
\tablefoot{The listed values are the median and median absolute deviation of the ratio between the reconstructed mean density, \nemean, and the true 3D density, \truene. Results are reported for masked and unmasked substructures and for non-parametric and parametric deprojection methods.}
\end{table}

\subsection{Dependence on cluster dynamical state}
Dynamically disturbed systems typically show larger biases due to the presence of substructures and deviations from spherical symmetry. These biases are reported in Table \ref{tab:density_bias_dynamic} and were extracted from the values presented in Fig. \ref{fig:morpho_profiles}.

\begin{table}[!h]
\caption{Dependence of density bias on dynamical state.}
\label{tab:density_bias_dynamic}
\centering
\resizebox{\columnwidth}{!}{
\begin{tabular}{lcccc}
\hline\hline
& \multicolumn{2}{c}{Most relaxed (MR)}
& \multicolumn{2}{c}{Most disturbed (MD)} \\
\cline{2-3}\cline{4-5}
Radius &
\nemedian/\truene &
\nemean/\truene &
\nemedian/\truene &
\nemean/\truene \\
\hline
$0.05\,R_{500}$ & $1.00 \pm 0.09$ & $1.01 \pm 0.10$ & $1.09 \pm 0.17$ & $1.10 \pm 0.16$   \\
$0.1\,R_{500}$ & $1.02 \pm 0.12$ & $1.04 \pm 0.13$ & $1.07 \pm 0.14$ & $1.12 \pm 0.13$   \\
$0.25\,R_{500}$ & $0.97 \pm 0.24$ & $1.01 \pm 0.23$ & $1.05 \pm 0.19$ & $1.08 \pm 0.16$   \\
$0.5\,R_{500}$ & $1.02 \pm 0.30$ & $1.13 \pm 0.31$ & $0.94 \pm 0.19$ & $1.12 \pm 0.22$  \\
$0.75\,R_{500}$ & $1.02 \pm 0.36$ & $1.08 \pm 0.40$ & $0.91 \pm 0.21$ & $1.08 \pm 0.19$   \\
$1.0\,R_{500}$ & $1.00 \pm 0.35$ & $1.01 \pm 0.36$ & $0.96 \pm 0.24$ & $1.10 \pm 0.19$  \\
$1.2\,R_{500}$ & $0.95 \pm 0.32$ & $1.06 \pm 0.32$ & $0.99 \pm 0.27$ & $1.11 \pm 0.19$  \\
\hline
\end{tabular}
}
\vspace{1mm}
\tablefoot{The values listed are the median and median absolute deviation of the reconstructed-to-true density ratios for the 20 MR and 20 MD clusters.}
\end{table}
\FloatBarrier
\subsection{Projection effects}
We report the intrinsic uncertainty introduced by projection effects 
drawn from the median of the profiles shown with the solid black line in Fig. \ref{fig:density_scatter} in Table \ref{tab:projection_scatter}.

\begin{table}[h]
\caption{Projection-induced scatter in the density profiles.}
\label{tab:projection_scatter}
\centering
\begin{tabular}{c c c}
\hline\hline
Radius & $\sigma$ \nemedian & $\sigma$ \nemean  \\
 &  &  \\
\hline
$0.05\,R_{500}$ & $0.06 \pm 0.02$ & $0.06 \pm 0.02$ \\
$0.1\,R_{500}$ & $0.06 \pm 0.02$ & $0.05 \pm 0.02$ \\
$0.25\,R_{500}$ & $0.06 \pm 0.02$ & $0.05 \pm 0.02$ \\
$0.5\,R_{500}$ & $0.06 \pm 0.02$ & $0.07 \pm 0.03$ \\
$0.75\,R_{500}$ & $0.06 \pm 0.02$ & $0.1 \pm 0.05$ \\
$1.0\,R_{500}$ & $0.06 \pm 0.02$ & $0.13 \pm 0.07$ \\
$1.2\,R_{500}$ & $0.06 \pm 0.02$ & $0.15 \pm 0.07$ \\
\hline
\end{tabular}
\tablefoot{The values listed are the median and median absolute deviation of the intrinsic scatter among the profiles reconstructed along 40 LOSs for the full sample.}
\end{table}
\subsection{Gas mass}

The gas mass within $R_{500}$ inherits the systematic behaviour of the density profiles used to derive it. Table \ref{tab:gas_mass_stats} reports the results of Sect. \ref{sec:gasmass}.

\begin{table}[h]
\caption{Reconstructed true gas mass ratios within $R_{500}$.}
\label{tab:gas_mass_stats}
\centering
\begin{tabular}{lc}
\hline\hline
\noalign{\smallskip}
Profile                  & Median Ratio  \\
                         & ($M_{\rm gas} / M_{\rm gas, 3D, 500}$)  \\
                         &  [\%] \\
\noalign{\smallskip}
\hline
\noalign{\smallskip}
 \nemedian         & $1.01 \pm 0.16$ \\
\nemean            & $1.10 \pm 0.18$ \\
\nemean\ MR        & $1.11 \pm 0.22$ \\
\nemean\ MR-HP     & $0.99 \pm 0.16$ \\
\noalign{\smallskip}
\hline
\end{tabular}
\tablefoot{The listed values are the median and median absolute deviation of the reconstructed-to-true gas mass ratio. MR denotes the most relaxed subsample, and MR-HP denotes the hand-picked relaxed subsample.}
\end{table}

\subsection{$d \log n_e$}

The logarithmic gradient of the density profile is accurately recovered outside the central region despite the method used to recover the density profiles. Table \ref{tab:slope_accuracy} summarises the results shown in Fig. \ref{fig:gradient_plot}.

\begin{table}[h]
\caption{Accuracy of logarithmic density gradients.}
\label{tab:slope_accuracy}
\centering
\begin{tabular}{c c c }
\hline\hline
Radius &   d $\log \widetilde{ n_{\rm e}}$ / d $\log$ \truene  & d $ \log \langle n_{\rm e} \rangle$ / d $\log$ \truene  \\
       &  [\%]                                                 & [\%]  \\
\hline
$0.05\,R_{500}$ & $1.28 \pm 0.37 $ & $1.27 \pm 0.28$ \\
$0.1\,R_{500}$ & $1.09 \pm 0.21 $ & $1.10 \pm 0.16$ \\
$0.25\,R_{500}$ & $1.02 \pm 0.16 $ & $0.99 \pm 0.13$ \\
$0.5\,R_{500}$ & $1.05 \pm 0.12 $ & $0.99 \pm 0.10$\\
$0.75\,R_{500}$ & $1.03 \pm 0.11 $ & $1. \pm 0.13$ \\
$1.0\,R_{500}$ & $0.98 \pm 0.11 $ & $0.96 \pm 0.15$\\
$1.2\,R_{500}$ & $1.00 \pm 0.12 $ & $1.00 \pm 0.14$\\
\hline
\end{tabular}
\tablefoot{The listed values are the median and median absolute deviation of the ratios between the reconstructed and true logarithmic density gradients at fixed radii.}
\end{table}

Both mean and median profiles reproduce the true logarithmic slope within $\sim5\%$ beyond $0.2R_{500}$.

\section{Mean versus median profiles: Other considerations}

\begin{figure}[!t]
\centering
\includegraphics[width=\columnwidth]{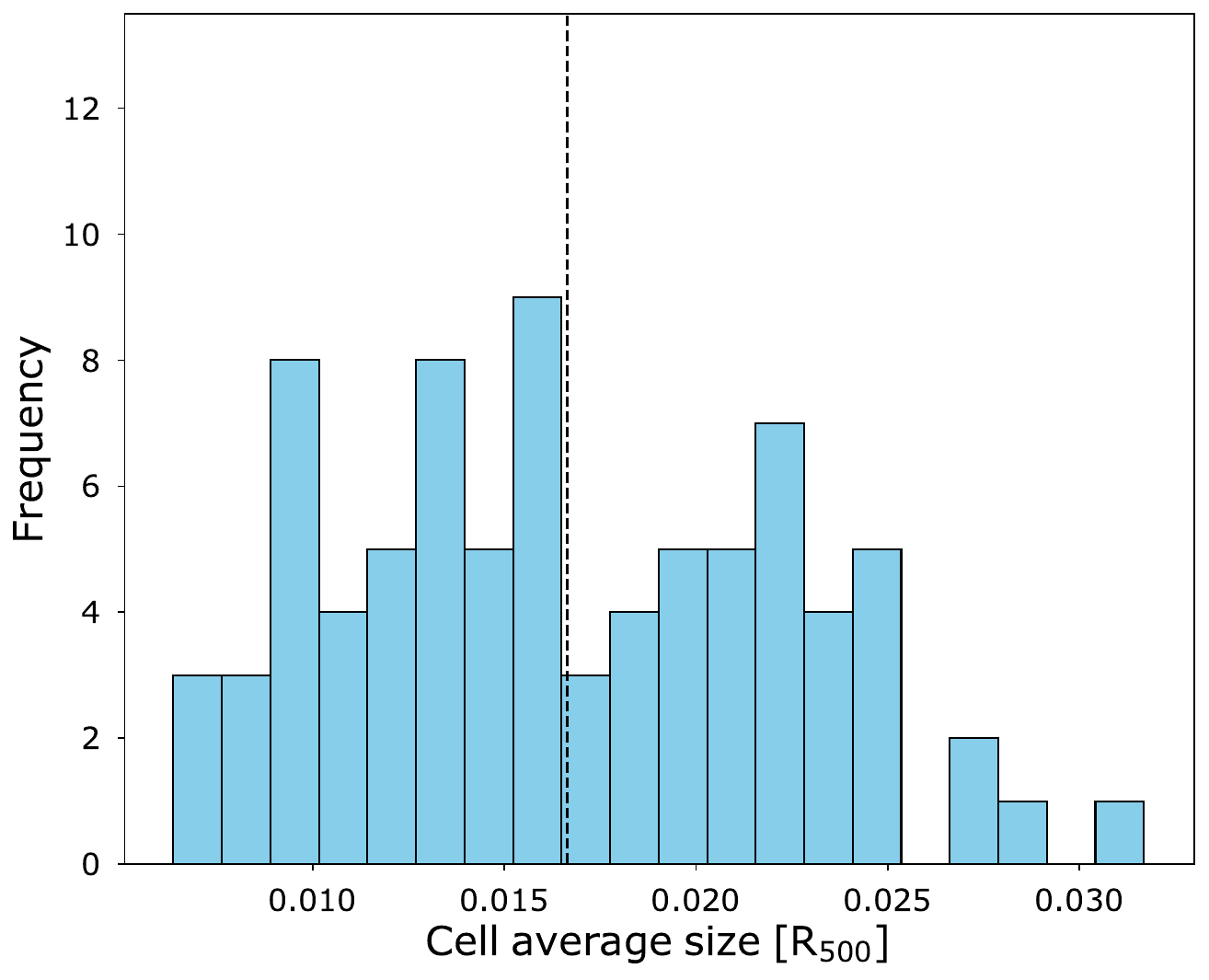}
\includegraphics[width=\columnwidth]{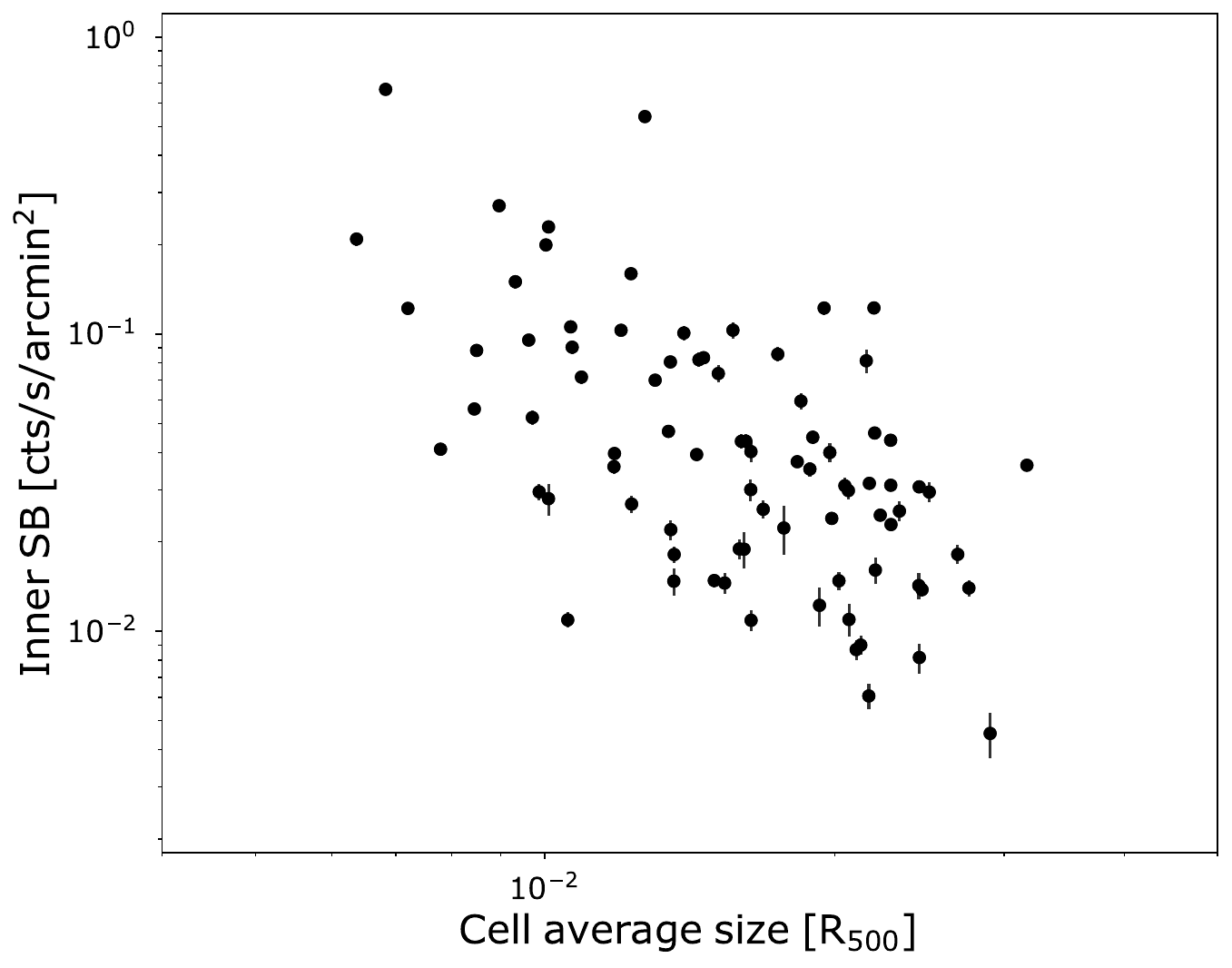}

\caption{Loss of precision due to Voronoi binning.
Top panel: Histogram of the average size of the Voronoi cells in terms of
\Rv\ within 0.1\Rv\ for all the clusters of the \chxmate\ collaboration.
The horizontal black line represents the average size of 0.017\Rv.
Bottom panel: Comparison between the average size of the Voronoi cells
and the value of the surface brightness in the inner bin.
The derivation of the Voronoi maps of the \chxmate\ collaboration and
the median surface brightness profiles are reported in
\citealt{2023Bartalucci}.}
\label{fig:voronoi_cell_size}
\end{figure}

The results shown in this work indicate that the density profile using median emission measure profiles is robust against local inhomogeneities and also the presence of substructures. However, we argue that both tools are fundamental to characterise the ICM properties in real observations. The median technique indeed is the ideal tool to derive quantities in the [0.1-1]\Rv\ range, but it requires a local estimation of the cluster surface brightness that is robust and statistically homogeneous. To satisfy this requirement, the use of Voronoi tessellation (\citealt{cappellari2003} and \citealt{diehl2006}) is essential. This adaptive binning method partitions the observational field into cells with a constant signal-to-noise ratio ensuring that each spatial bin carries a comparable amount of statistical information. This step is necessary for the application of the median by providing a uniform statistical weight to each cell. However, in observations the use of Voronoi tessellation to compute median profiles reduces the effective spatial resolution. Moreover, deviations from spherical symmetry can also affect the relation between mean and median profiles, complicating the interpretation of clumping measurements.

We used the Voronoi binned maps derived for the \chxmate\ collaboration \citep{2023Bartalucci} and determined the size of the cells within the inner region at R<0.1\Rv. The histogram average size of the cells is reported in the top panel of Fig. \ref{fig:voronoi_cell_size} and the average value of the cell, $\langle Size \rangle = 0.017$\Rv, is indicated with the horizontal dotted line. This indicates that, on average, the surface brightness inner radial bin cannot probe smaller regions. However, the mean density profiles derived in \citealt{2024Riva} (cfr. Fig. C2) have been used by the authors to measure the entropy profiles below 0.01\Rv. Furthermore, the average size of the cell depends on the cluster central brightness. The bottom panel of Fig. \ref{fig:voronoi_cell_size} shows the anti-correlation between the Voronoi cell size and the inner radial bin brightness. 

We advocate that median and mean profiles are both fundamental tools to derive the ICM density profiles and needs to be used depending on the science case. For this reason, the \chxmate\ collaboration is providing surface brightness and the derived deprojected quantities using both type of profiles.
\end{appendix}
\end{document}